\documentclass{ieeeaccess}
\usepackage{cite}
\usepackage{amsmath,amssymb,amsfonts}
\usepackage{algorithm} 
\usepackage{algpseudocode} 
\usepackage{url}
\definecolor{lightgray}{gray}{0.9} 
\usepackage{tabularx}
\usepackage[dvipsnames]{xcolor}
\usepackage{booktabs}
\usepackage[colorlinks=true,linkcolor=blue,anchorcolor=black,citecolor=teal,filecolor=black,menucolor=black,runcolor=black,urlcolor=black]{hyperref}
\usepackage{soul}

\usepackage{array}
\usepackage{comment}
\usepackage{fourier} 
\usepackage{makecell}
\usepackage[normalem]{ulem}
\sethlcolor{yellow}

\usepackage{etoolbox}
\usepackage{longtable}
\usepackage{graphicx}
\usepackage{lipsum}
\usepackage{float}
\usepackage{multicol}
\usepackage{multirow}
\usepackage{textcomp}
\usepackage{subcaption}
\usepackage{multirow}
\usepackage{mdframed}
\mdfdefinestyle{custombox}{
    linecolor=blue,        
    linewidth=2pt,         
    roundcorner=10pt,      
    backgroundcolor=blue!5, 
    innertopmargin=10pt,
    innerbottommargin=10pt,
    innerleftmargin=10pt,
    innerrightmargin=10pt
}
\usepackage{etoolbox} 

\definecolor{grayheader}{RGB}{0, 0, 139}
\definecolor{cadetblue}{RGB}{139, 0, 0}

\newmdenv[
  linewidth=1.2pt,
  linecolor=grayheader,
  roundcorner=6pt,
  backgroundcolor=white,
  innertopmargin=1em,
  innerbottommargin=1em,
  innerleftmargin=1em,
  innerrightmargin=1em,
  frametitlebackgroundcolor=grayheader,
  frametitlefont=\color{white}\bfseries,
  frametitle=Prompt
]{promptbox}

\newmdenv[
  linewidth=1.2pt,
  linecolor=cadetblue,
  roundcorner=6pt,
  backgroundcolor=white,
  innertopmargin=1em,
  innerbottommargin=1em,
  innerleftmargin=1em,
  innerrightmargin=1em,
  frametitlebackgroundcolor=cadetblue,
  frametitlefont=\color{white}\bfseries,
  frametitle=Response
]{responsebox}
\def\BibTeX{{\rm B\kern-.05em{\sc i\kern-.025em b}\kern-.08em
    T\kern-.1667em\lower.7ex\hbox{E}\kern-.125emX}}    
\begin{document}
\history{Date of publication xxxx 00, 0000, date of current version xxxx 00, 0000.}
\doi{10.1109/ACCESS.2024.DOI}

\title{Large Language Models for Power System Security: A Novel Multi-Modal Approach for Anomaly Detection in Energy Management Systems}
\author{\uppercase{Aydin Zaboli}\authorrefmark{1}, \IEEEmembership{Member, IEEE}, 
\uppercase{Junho Hong}\authorrefmark{1}, \IEEEmembership{Senior Member, IEEE},
\uppercase{Alexandru Ștefanov}\authorrefmark{2}, \IEEEmembership{Member, IEEE}, 
\uppercase{Chen-Ching Liu}\authorrefmark{3}, \IEEEmembership{Life Fellow, IEEE},
\uppercase{Chul-Sang Hwang}\authorrefmark{4}, \IEEEmembership{Member, IEEE}}

\address[1]{Department of Electrical and Computer Engineering, University of Michigan -- Dearborn, MI, 48128 USA.}
\address[2]{Department of Electrical Sustainable Energy, Technische Universiteit Delft, 2628 CD Delft, Netherlands.}
\address[3]{Bradley Department of Electrical and Computer Engineering, Virginia Polytechnic Institute and State University, Blacksburg, VA 24061, USA.}
\address[4]{Smart Grid Research Division System Reliability Research Team, Korea Electrotechnology Research Institute (KERI), Gwangju-si, 61751, South Korea.}


\tfootnote{This work was supported by the Korea Institute of Energy Technology Evaluation and Planning (KETEP) grant funded by the Ministry of Trade, Industry and Energy (MOTIE) of the Republic of Korea (RS2022-KP002883, Building Crucial Infrastructure in order for Demonstration Complex Regarding Distributed Renewable Energy System).}

\markboth
{Author \headeretal: Preparation of Papers for IEEE TRANSACTIONS and JOURNALS}
{Author \headeretal: Preparation of Papers for IEEE TRANSACTIONS and JOURNALS}

\corresp{Corresponding author: Chul-Sang Hwang (e-mail: hcs1006@keri.re.kr).}

\begin{abstract}
This paper elaborates on an extensive security framework specifically designed for energy management systems (EMSs), which effectively tackles the dynamic environment of cybersecurity vulnerabilities and/or system problems (SPs), accomplished through the incorporation of novel methodologies. A comprehensive multi-point attack/error model is initially proposed to systematically identify vulnerabilities throughout the entire EMS data processing pipeline, including post state estimation (SE) stealth attacks, EMS database manipulation, and human-machine interface (HMI) display corruption according to the real-time database (RTDB) storage. This framework acknowledges the interconnected nature of modern attack vectors, which utilize various phases of supervisory control and data acquisition (SCADA) data flow. Then, generative artificial intelligence (GenAI)-based anomaly detection systems (ADSs) for EMSs are proposed for the first time in the power system domain to handle the scenarios. Further, a set-of-mark generative intelligence (SoM-GI) framework, which leverages multimodal analysis by integrating visual markers with rules considering the GenAI capabilities, is suggested to overcome inherent spatial reasoning limitations. The SoM-GI methodology employs systematic visual indicators to enable accurate interpretation of segmented HMI displays and detect visual anomalies that numerical methods fail to identify. Validation on the IEEE 14-Bus system shows the framework's effectiveness across scenarios, while visual analysis identifies inconsistencies. This integrated approach combines numerical analysis with visual pattern recognition and linguistic rules to protect against cyber threats and system errors.
\end{abstract}


\titlepgskip=-15pt

\maketitle
\section{Introduction}
\PARstart{T}he accelerated digitization of power systems has revolutionized EMSs into complex cyber-physical architectures capable of coordinating electricity delivery across expansive grids. Modern EMS frameworks integrate SCADA systems with cutting-edge computational algorithms, promoting a complex network in which field equipment such as remote terminal units (RTUs) and phasor measurement units (PMUs) transmit synchronized, real-time data that enables enhanced monitoring and control of grid performance~\cite{sun2018cyber, gungor2011smart, hong2022automated}. While this integration brings substantial operational efficiencies and enhanced operational visibility, it also expands the system’s vulnerability footprint, exposing multiple attack vectors within the data acquisition and processing sequence. The convergence of operational technology (OT) and information technology (IT) domains necessitates a reconsideration of traditional cybersecurity approaches. 
SE is crucial in power systems, offering operators a real-time overview of the system's condition to maintain reliability and control~\cite{li2020online}. However, the coupling of power systems with cyber infrastructure increases their exposure to complex cyber-attacks. Among these, false data injection (FDI) attacks present a particularly serious threat by manipulating measurement data to lead system operators toward incorrect control actions while evading traditional bad data detection (BDD) methods~\cite{liu2011false}. Also, stealth attacks represent a sophisticated subset of FDI attacks. These attacks are mathematically designed to bypass traditional BDD mechanisms by ensuring that the measurement residuals remain practically unchanged from normal conditions~\cite{liang2017review}. Hence, BDD techniques such as $\chi^2$-test, primarily based on SE residuals, are increasingly insufficient, particularly against sophisticated stealth attacks and FDI strategies. 
These stealthy attacks can manipulate PMU and SCADA data while maintaining consistency with system models, evading conventional BDD filters~\cite{paudel2018stealthy, jiang2023monitoring}. In parallel, attackers might manipulate data within HMIs to introduce misleading display changes that mislead operators, all the while not necessitating the modification of actual physical measurements. Given the critical infrastructure, modern power grids necessitate robust mechanisms capable of identifying both mathematically stealthy manipulations and operator-level display tampering. Recent literature emphasizes hybrid detection approaches that synthesize physical-model awareness, data-driven analytics, and temporal behavioral models to detect anomalies across multi-layered attack vectors~\cite{motakatla2023cybersecurity}. In accordance with this, testbed-based experiments and co-simulation studies emphasize the importance of AI and multi-modal detection approaches to limit false positives (FPs) while improving sensitivity to complex attacks~\cite{choi2024generative, zaboli2025advanced, zaboli2024leveraging}.
\subsection{Problem Statement}
The evolving cybersecurity threats targeting EMSs represent a new class of challenges that exceed the capabilities of traditional defenses originally designed for isolated OT domains~\cite{choi2024generative}. As EMS architectures increasingly integrate both cyber and physical layers, malicious actors are now capable of executing coordinated attacks that span the entire data processing pipeline, from the collection of measurements through the SE stage and finally to the HMI screen. An alarming feature of such breaches is their ability to maintain statistical validity, enabling them to evade conventional BDD techniques such as those relying on $\chi^2$ hypothesis testing. These attack pathways are diverse and sophisticated, such that some are stealth attacks or may involve tampering with SE outputs after validation but before storage in historical databases, effectively corrupting trusted data storage systems. Others may directly target the EMS databases, compromising both operational and historical data integrity. Also, some attacks on HMIs can occur, which can subtly manipulate visual representations such as circuit breaker (CB) statuses or voltage levels without modifying the core numerical calculations, thereby misleading operators and delaying response times despite accurate SE outputs~\cite{choi2024generative}. The complexity of power systems, intensified by the integration of distributed energy resources (DERs), further increases these vulnerabilities. Within such dynamic conditions, malicious data injections can be designed to imitate normal operations, potentially leading the system into inefficient or unstable operating states without timely identification.

Also, traditional anomaly detection (AD) mechanisms, such as the SE process, focus on numerical inconsistencies and threshold-based alarms, which are often insufficient when attackers manipulate the visual elements of SCADA displays. For instance, falsified visual indicators may not breach statistical thresholds but can still mislead human operators, especially during high-stress operational conditions. Although recent advances in GenAI offer promising opportunities for pattern recognition and AD processes, they present distinctive challenges. Many generative models still lack robust spatial reasoning capabilities and often require explicit guidance or structured prompts to accurately interpret the content of segmented or context-rich HMI displays. The absence of integrated detection frameworks that simultaneously assess numerical validity, visual consistency, and semantic rule coherence creates substantial gaps in current EMS cybersecurity strategies. These gaps provide opportunities for sophisticated attackers to exploit the interfaces between detection layers and human perception, particularly in scenarios requiring rapid and confident operator decision-making. Addressing these vulnerabilities requires a framework shift toward multi-modal, intelligent security solutions capable of bridging these aspects of grid operations~\cite{sun2025data}.
\subsection{Research Objectives}
This section proposes the design of a multi-layered AD framework tailored for EMSs, aiming to address a broad range of security vulnerabilities by combining advanced computational techniques with domain-specific operational insights. The central goal is to construct and evaluate a multi-point detection strategy that continuously observes critical stages across the EMS data flow, focusing in particular on post-SE validation based on stealth attacks, database integrity verification, and the validation of HMI outputs while ensuring the efficiency necessary for grid stability and operational continuity. A key innovation of this research lies in the development of a SoM-GI approach that is designed to overcome spatial reasoning constraints often encountered in models. By embedding structured visual indicators, directional symbols, and connection point (CP) annotations within screen segments, the proposed method enhances the interpretability of complex HMI layouts and facilitates the identification of hidden anomalies. The effectiveness of the proposed framework will be assessed through testing on an IEEE 14-bus system that meets the North American Electric Reliability Corporation (NERC) regulations in terms of voltage violations. Validation efforts will include detection scenarios involving manipulated state vectors, fabricated topology information, and HMI RE-based deception.
\subsection{Related Work}
Generative pre-trained transformers (GPTs) can enhance the system diagnosis (SD) accuracy beyond traditional machine learning (ML) and deep learning (DL) models and BDDs through contextual processing and adaptability. Unlike ML models requiring extensive pattern training, GPTs can simultaneously analyze data, historical patterns, and operator inputs. Their NLP capabilities enable the integration of operator logs with numerical data, leading to more accurate SD processes in EMSs. They can also adapt to new system configurations without complete retraining, making them more efficient for evolving power network topologies while maintaining high diagnostic accuracy during critical situations~\cite{zaboli2024chatgpt, zhang2025review, cetinay2017comparing, NREL_87740}. Furthermore, it is challenging for BDDs to detect stealth attacks as they evade detectors as well as unexpected scenarios. Thus, Table~\ref{tab:literature_survey_EMS_stealth} provides a concise review of relevant studies.
\begin{table*}[!h] 
\centering
\caption{A literature survey on the AD process in EMSs.}
\label{tab:literature_survey_EMS_stealth}
\small
\begin{tabular}{|p{4.5cm}|p{5.75cm}|p{5.75cm}|}
\hline
\textbf{Author} & \textbf{Contributions} & \textbf{Challenges} \\
\hline
Falconer~\textit{et al.}~\cite{falconer2022leveraging} (2022) & 
\textbullet~ML-based approximation of complex power flow (PF) problems\newline
\textbullet~Addressed unit commitment and security restrictions &
\textbullet~Scalability issues with fully connected networks\newline
\textbullet~Restricted convolutional neural network (CNN) accuracy in anomaly scenarios \\
\hline
Mukherjee~\cite{mukherjee2022novel} (2022) & 
\textbullet~Multi-label classification framework for FDI diagnosis\newline
\textbullet~Model-free detection without grid/attack knowledge &
\textbullet~Limited comprehensiveness for diagnosing unexpected anomalies\newline
\textbullet~Challenges in retraining models on new attacks \\
\hline
Ashrafuzzaman~\textit{et al.}~\cite{ashrafuzzaman2020detecting} (2020) & 
\textbullet~Ensemble ML-based detection of stealthy FDIs\newline
\textbullet~Inclusion of Random Forest Classifier for feature reduction &
\textbullet~Extensive labeled data requirement\newline
\textbullet~High FP rates in unsupervised scenarios \\
\hline
Guo~\textit{et al.}~\cite{guo2023event} (2023) & 
\textbullet~Event-driven FDI attack strategy\newline
\textbullet~Real-time residual-driven attack scheduling &
\textbullet~Dependence on accurate real-time residual calculations\newline
\textbullet~Performance reduction in noisy conditions \\
\hline
Zhang~\textit{et al.}~\cite{zhang2023limitation} (2024) & 
\textbullet~Comprehensive analysis of reactance perturbation strategy limitations\newline
\textbullet~Proposed enhanced reactance perturbation strategy &
\textbullet~Need for detailed topology analysis\newline
\textbullet~Scalability issues for larger systems \\
\hline
Guo~\textit{et al.}~\cite{guo2023residual} (2023) & 
\textbullet~Optimal residual-based FDI for multi-sensor systems\newline
\textbullet~Sensor selection principle to maximize deterioration &
\textbullet~Computational complexity of optimization\newline
\textbullet~High resource demand for real-time applications \\
\hline
\end{tabular}
\end{table*}
Ashrafuzzaman \textit{et al.}~\cite{ashrafuzzaman2020detecting} introduced a data-driven ensemble ML framework aimed at the detection of stealthy FDI attacks within smart grids by employing classification algorithms. Although this approach enhances the AD process by mitigating the complexities associated with high-dimensional data, it faces significant challenges, particularly the requirement for extensive labeled datasets and the frequency of high FP rates in unsupervised scenarios. An approach addressing stealth sensor and actuator attacks under resource constraints on discrete event systems using supervisory control was developed by He~\textit{et al.}~\cite{he2025cyber}. They introduced combined and efficient vulnerability and algorithmic techniques. However, their approach is constrained by the computational complexity of modeling extensive discrete event system scenarios. Guo~\textit{et al.}~\cite{guo2023event} introduced an event-driven stealthy FDI attack strategy against remote SE systems. Their method dynamically initiates attacks based on real-time residuals to deteriorate system performance optimally. Despite its effectiveness in resource-limited contexts, it depends on accurate real-time residual computations, which could be challenging in noisy environments. 

Lee~\textit{et al.}~\cite{IJAMD01624} proposed a unified industrial large knowledge model framework, emphasizing how domain-specific knowledge can be integrated with large language models (LLMs) to address complex industrial challenges. Their framework introduces the concept of combining human-interpretable data with structured machine-generated data through LLM-based semantic understanding. The authors demonstrated that LLMs can serve as domain experts by processing diverse industrial data and generating actionable insights, though they noted challenges in reliability and privacy that require specialized training approaches. Building on industrial LLM applications, Aberbach~\textit{et al.}~\cite{aberbach2025ai} investigated how LLMs can enhance smart grid resilience and efficiency by processing unstructured text-based information alongside traditional grid data. Their work highlighted that LLMs excel at handling diverse data types including operator queries, maintenance logs, and regulatory documents. They emphasized the importance of combining LLMs with specialized algorithms for grid-specific tasks, suggesting a hybrid approach similar to the integration of prompt engineering with meta-learning techniques, without multimodal analysis considering the analysis of HMI screen and visual interpretation of power system components. Further, Li~\textit{et al.}~\cite{li2025g} introduced a graph transformer-based vision-language model for industrial AD processes. Their research demonstrated that aligning visual features with textual descriptions through prompt engineering significantly improves AD accuracy across different product categories. The study showed that multi-level domain adaptation enables models to capture semantic consistency of anomalous features at various scales, achieving strong performance even with limited training data. However, their approach focused primarily on visual anomalies in manufacturing rather than cyber-physical attacks in smart grids. More directly related to grid security, Shen~\textit{et al.}~\cite{shen2025large} developed an LLM-based security situation awareness framework specifically for smart grids. Their approach transformed diverse grid data into structured text prompts and utilized LLM reasoning capabilities to detect threats and forecast security states. The research demonstrated that LLMs could process electrical parameters, meteorological data, and social factors in a unified semantic space, achieving improved prediction accuracy through multimodal fusion. Nevertheless, their work concentrated on general security awareness rather than addressing the specific challenge of zero-day attack detection in federated environments. Addressing privacy concerns in distributed grid systems, Dasgupta and Mitra~\cite{dasgupta2025large} proposed a federated zero-shot learning framework that leverages LLMs to generate client-specific semantic embeddings for intrusion detection. Their approach demonstrated how LLMs could create varied textual descriptions of attacks for each client, reducing the risk of inference attacks while maintaining detection accuracy. By combining federated learning with zero-shot capabilities, they showed that models could detect unknown attacks without sharing sensitive grid data. However, their methodology differs from the proposed approach in that they focused specifically on intrusion detection for network-based attacks rather than the broader operational AD and forecasting tasks considering the visual interpretation and power flow modeling.
A perturbation strategy for defending against FDI attacks in IoT-based smart grids was presented by Zhang~\textit{et al.}~\cite{zhang2023limitation}. They demonstrated theoretically and numerically that inappropriate selection of branches for reactance perturbation could compromise defense effectiveness. However, their enhanced strategy requires detailed prior topology analysis, potentially limiting its scalability to larger systems. Guo~\textit{et al.}~\cite{guo2023residual} introduced a residual-based stealthy FDI attack for multi-sensor estimation systems, highlighting the critical selection of sensors under resource constraints to optimize degradation. Their approach leverages historical and current residuals to enhance attack impact but demands significant computational resources to solve optimization problems at each step. Zhou~\textit{et al.}~\cite{zhou2025worst} investigated optimal FDI attacks against partially secured remote SE systems by formulating optimization problems to maximize SE errors. They developed strong detection and estimation strategies, but real-world complexity and the need for secure communication may impact practicality.
A framework utilizing ML algorithms to approximate the PF analysis was developed by Falconer~\textit{et al.}~\cite{falconer2022leveraging}. This approach supported the resolution of complex PF problems that incorporated unit commitment and security restrictions. However, their model faces challenges stemming from scalability issues associated with fully connected networks as the system size increases. Additionally, CNN models demonstrate restricted predictive accuracy due to their dependence on convolutions if there are anomalies. 
Hu~\textit{et al.}~\cite{hu2024anomaly} introduced the state deviation index for the diagnosis of FDI attacks and sudden load changes, demonstrating its effectiveness in IEEE 14-bus and 30-bus systems. However, the algorithm exhibited limited adaptability when faced with different errors. This suggests the need for an approach in which indicators are carefully designed to reflect unique properties of specific anomalies. The absence of such tailoring risks compromising the accuracy of evaluation metrics, presenting significant FP and false negative (FN) results. A multi-variable long short-term memory autoencoder (LSTM-AE) had been formulated for an SD process by Sarker~\textit{et al.}~\cite{10302416}. Their model successfully diagnosed SPs associated with errors, validated through a 123-bus unbalanced distribution network. However, according to sophisticated mathematical modeling of the PF analysis considering the SD process, developing the model with different abnormal scenarios in ML algorithms is challenging. Since they jeopardize the accuracy of the proposed algorithm, which can be time-consuming, they need more effort. Furthermore, a graphical user interface (GUI) of an EMS could show abnormal information due to system errors or bugs. These errors are not easy for SCADA control room engineers to diagnose due to the huge volume of information. These issues cannot be managed by the SE process and need adaptive solutions. A novel FDI diagnostic method using LSTM-AE and CNN-AE with an unsupervised learning approach was introduced in~\cite{ganjkhani2024application}, avoiding the need for anomalous data during training. It also proposed an LSTM variational AE-based reconstruction method to maintain stability by closely replicating the original data from anomalous data. Despite this, the reliance on unsupervised learning presents challenges in certain scenarios. Hence, zero-day attacks need retraining of ML algorithms, which is time-intensive. Also, collecting all unknown errors can take much effort~\cite{guarino2023two, xue2024real}. Mukherjee~\cite{mukherjee2022novel} introduced an approach to identify FDI attacks using a multi-label classification framework. This method leveraged conventional bad-data detectors to enhance measurement accuracy and diagnose unstructured attacks. Accordingly, this model-free strategy required no prior knowledge of grid or attack vectors, making it a highly effective solution for the FDI diagnosis. Nevertheless, a consideration of an FDI attack cannot merely show the comprehensiveness of this algorithm. Also, there are some malfunctions that cannot be properly diagnosed by ML models. A retraining on new attacks and the diagnosis of other malfunctions (e.g., a CB can be opened during a fault in the normal operation; however, this could be because of inaccurate communication that sends the status of the CB to the SCADA room) are challenging for ML techniques~\cite{faramondi2023evaluating, rakas2020review, buslon2023attack}.
Although recent studies have achieved notable advancements in utilizing LLMs for smart grids, there are still numerous areas that have not been fully addressed. Current methodologies either emphasize broad industrial frameworks that neglect grid-specific enhancements, concentrate solely on AD without predictive functions, or manage privacy with federated learning yet fail to achieve the necessary multimodal integration for robust smart grid security.

\subsection{Contributions}
The integration of GenAI tools with AD techniques holds significant promise for revolutionizing the SE process and PF analysis. By leveraging the capabilities of GPT tools (e.g. Anthropic Claude Pro~\cite{anthropic}) to understand and interpret the natural language processing (NLP) of snapshots and network displays in addition to data analysis and understanding of the mathematical modeling, this approach enables a more robust diagnosis of anomalies/errors within the SE process and PF information. The combination of NLP and visual analytic techniques leads to advancements in the SD process, the development of user-friendly interfaces for power system monitoring, and a simplification of diagnostic processes within the energy sector, particularly in cases of unknown errors/attacks in the visual information. To tackle these gaps, two significant contributions are presented, advancing the state-of-the-art in the AD process for EMSs, according to the literature surveys presented in the previous section as follows:
\begin{itemize}
    \item \textbf{Multi-Point Attack Detection Framework along with GenAI-based AD Processes:} A GenAI-driven ADS model has been innovatively designed to transform the detection and resolution of vulnerabilities in the EMS data processing pipeline, targeting multi-point attacks and errors. In contrast to conventional approaches that focus on isolated detection points, this framework recognizes that sophisticated cyber-attacks exploit multiple stages of the SCADA data flow. The model specifically targets three critical vulnerability points including stealth attacks which can evade the BDDs, EMS database manipulation incidents (particularly FDI attacks following the SE process), and the HMI display corruption by manipulating the RTDB. 
    \item \textbf{Generative Intelligence-Enhanced Multimodal Analysis Framework:} Addressing the integration challenges and anomaly identification requirements, an innovative SoM-GI framework is proposed. This contribution bridges the gap between traditional numerical SE techniques and optimal PF results and emerging visual AD capabilities. The framework exploits GenAI's capabilities, augmented by visual markers and indicators alongside rules, to facilitate concurrent analysis of image and text data in the EMS environment.
    The SoM-GI methodology overcomes inherent spatial reasoning limitations in current GenAI systems by implementing systematic visual indicators, including CB status markers, directional transmission line indicators, and CP identifiers. These markers guide the AI's interpretation of segmented HMI displays, enabling accurate detection of visual anomalies inconsistencies and falsifications that often bypass numerical detection methods. Validation results confirm the framework's ability to identify sophisticated attacks that manipulate display segments while maintaining consistency in core data structures. This novel approach establishes a new paradigm for comprehensive security monitoring in power systems, integrating visual pattern recognition with linguistic rule processing. By standardizing the fusion of multimodal analysis techniques, the framework ensures that visual-based ADSs can be incorporated into existing EMS architectures without disrupting critical operational processes, thereby providing a practical solution.
\end{itemize}
\subsection{GenAI vs. Traditional ML/DL Approaches for Anomaly Detection in EMSs} \label{sec:ml_vs_genai}
The literature on the AD process in EMS domains has extensively explored classical ML and DL techniques. For example, unsupervised autoencoder frameworks reconstruct normal behavior and flag deviations via reconstruction error which in one recent industrial-substation case, the authors leveraged an LSTM-autoencoder and demonstrated that anomalies in remote-terminal sensing can be identified via elevated reconstruction losses~\cite{shrestha2024anomaly}. In another study, graph neural network (GNN)-based methods contextualize measurement correlation across network nodes and show improved localization of cyber-injections compared to flat autoencoders~\cite{yin2024advancing}. Moreover, as reviewed in~\cite{banik2023anomaly}, many smart grid AD methods still rely on large volumes of labeled anomalies, assume stationary behavior, and are less capable of rapidly adapting to new attack modes.

In contrast, the proposed GenAI framework introduces several key differentiators. First, whereas conventional DL approaches often learn fixed mappings from feature space to latent representations (e.g., encoder–decoder), this generative model utilizes LLM-style embeddings and interactions to capture dynamic anomaly signatures and subtle deviations from normal. Second, by integrating a human-in-the-loop (HITL) feedback pathway, the system can incrementally refine its prompt embeddings in response to newly observed patterns, reducing the need for extensive labeled anomaly datasets. Third, this framework addresses zero-day or novel attack vectors using the generative synthesis of anomalous scenarios and continuous adaptation of the model’s understanding of normal versus abnormal behavior. Together, these capabilities position the GenAI approach as more flexible and future-proof for EMS/SCADA AD processes than traditional ML/DL baselines. Table~\ref{tab:genai_vs_traditional} provides a systematic comparison of these approaches across key dimensions relevant to the EMS AD process, illustrating why GenAI represents a fundamental transformation rather than merely an incremental improvement over existing methodologies.

\begin{table*}[!h]
\caption{Comparison of GenAI vs. Traditional ML/DL Approaches for EMS Anomaly Detection~\cite{banik2023anomaly}.}
\label{tab:genai_vs_traditional}
\centering
\begin{tabular}{|p{3cm}|p{5cm}|p{5cm}|}
\hline
\textbf{Characteristic} & \textbf{Traditional ML/DL} & \textbf{Proposed GenAI Framework} \\
\hline
\textbf{Training Data Requirements} & Requires extensive labeled datasets covering all attack scenarios and operational conditions & Leverages pre-trained knowledge; minimal domain-specific training data needed. \\
\hline
\textbf{Feature Engineering} & Manual feature selection and engineering required; domain expertise intensive & Automatic feature extraction from multimodal inputs; minimal manual engineering. \\
\hline
\textbf{Adaptability} & Limited; requires complete retraining for new configurations or attack patterns & High; few-shot and zero-shot learning enable rapid adaptation without retraining \\
\hline
\textbf{Modality Processing} & Typically single-modality (numerical OR visual, not both) & Native multimodal processing (numerical, visual, textual simultaneously) \\
\hline
\textbf{Semantic Understanding} & Pattern recognition without semantic grounding in physical principles & Deep semantic understanding of power system physics and operational constraints \\
\hline
\textbf{Explainability} & Limited; black-box models with difficult-to-interpret outputs & High; natural language explanations with specific rule/constraint violations cited \\
\hline
\textbf{Zero-Day Attack Detection} & Poor; can only detect patterns seen during training & Strong; can reason about novel attack patterns using physical principles \\
\hline
\textbf{Domain Knowledge Integration} & Difficult; requires encoding expert knowledge into model architectures & Natural; incorporates domain knowledge through natural language descriptions \\
\hline
\textbf{Deployment Time} & Months (data collection, labeling, training, validation) & Days to weeks (prompt engineering, rule specification, testing) \\
\hline
\textbf{Visual Anomaly Detection} & Limited to computer vision models without operational context & Integrated visual-semantic analysis with SoM-GI methodology \\
\hline
\end{tabular}
\end{table*}

It is important to emphasize that GenAI does not necessarily replace traditional ML/DL approaches but rather complements them in a layered defense strategy. Conventional ML methods remain valuable for high-frequency, low-latency detection tasks where millisecond response times are critical. However, for complex reasoning tasks that require integration of multiple information sources, validation against physical constraints, interpretation of visual displays, and explanation of findings to human operators, GenAI offers capabilities that traditional approaches cannot match. This comparative analysis establishes the distinctive value proposition of the proposed GenAI-based framework and clarifies why it represents a significant advancement in EMS AD capabilities beyond what traditional ML/DL approaches can provide.
\subsection{Paper Structure}
The rest of this paper is organized as follows: Section~\ref{model-section} presents a multi-point attack model considering the EMS workflow, before and after the SE process. Section~\ref{implementation-section} demonstrates the comprehensive description of different attack points, their modeling, miscellaneous scenarios, and the GPT implementation with some direct responses using the trained GPT model within the power system domain. Finally, conclusions and directions for future work are outlined in Section~\ref{conclusion-section}.
\section{A Proposed Multi-point Attack Model in Energy Management Systems} \label{model-section}
\begin{figure}[!h]
\centerline{\includegraphics[width=1.0\columnwidth]{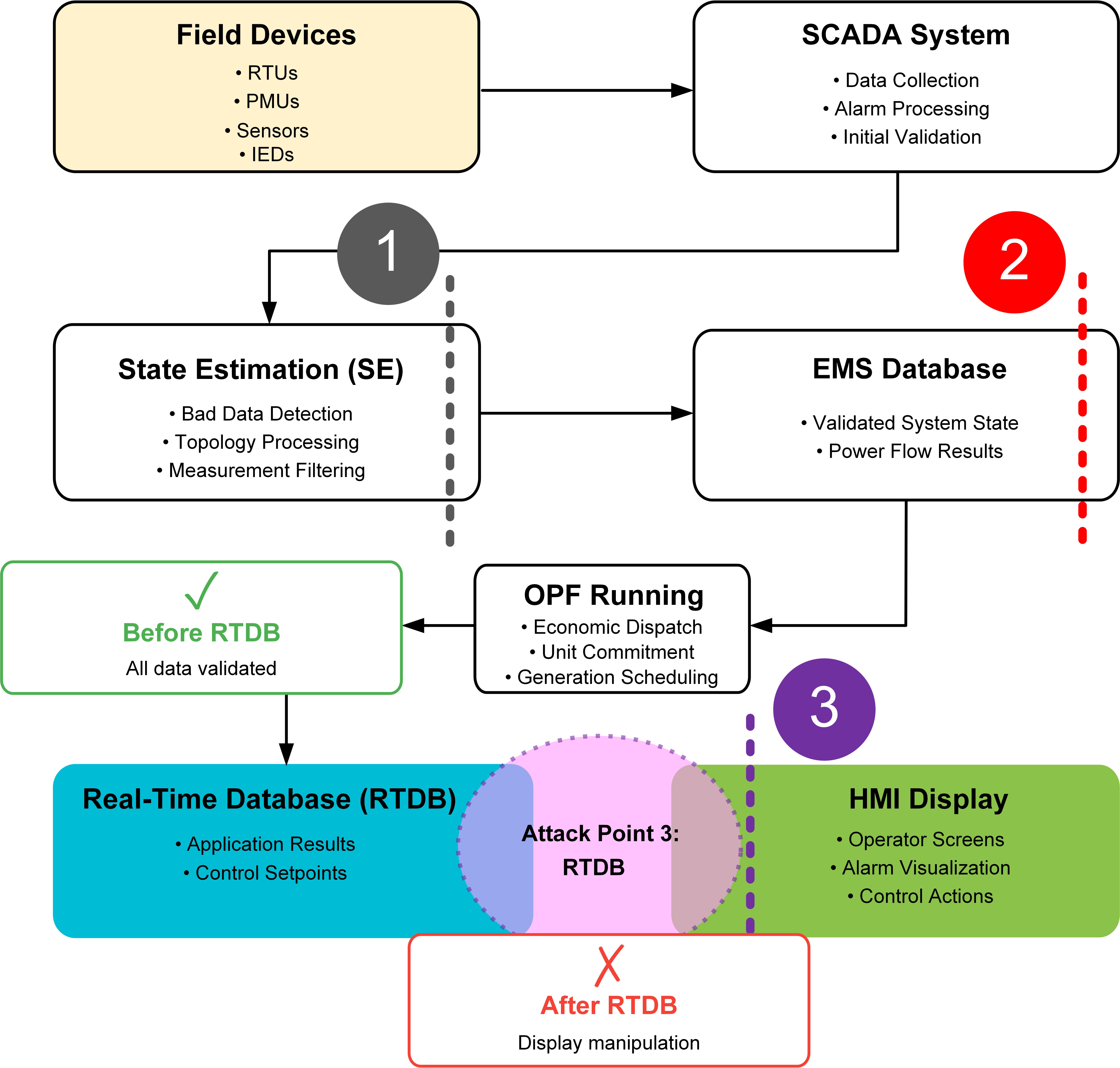}}
\caption{A general proposed framework for different attack points in EMSs.}
\label{attack_model_v4}
\end{figure}

Different attack points based on stealth attacks (i.e., attack point \#1) and intentional/unintentional attacks (i.e., attack points \#2 and \#3) including cyberattacks, system errors, and FDI attacks, are represented in Fig.~\ref{attack_model_v4}. 
It is evident that a variety of attacks/errors can appear at different points ranging from field devices to the HMI Display section within an EMS. The field devices represent the sensory layer of the SCADA architecture, functioning as the principal interface between the physical infrastructure and the digital control system. RTUs serve as industrial computer systems interfacing directly with physical equipment, conducting the critical role of converting analog signals from field devices into digital data while concurrently executing control commands transmitted from the SCADA system. These units are deployed in industrial settings and are required to maintain dependable operation under extreme conditions~\cite{pliatsios2020survey, menzel2021securing}. PMUs are advanced field devices that provide precise electrical waveform measurements. They deliver synchronized, real-time voltage and current phasor data for extensive regional monitoring and dynamic grid system analysis. The sensor network comprises a wide range of sensors that constantly monitor essential parameters such as flow rates and voltage levels, alongside the operational status of components. Additionally, intelligent electronic devices (IEDs), microprocessor-driven controllers, execute complex functions for protection, control, and monitoring of power equipment, often making autonomous decisions to improve system resilience~\cite{cheng2023survey}. The SCADA system serves as the central hub for data coordination and initial processing in industrial operations. It continuously collects real-time operational data from field devices using protocols such as DNP3, IEC61850, and Modbus, with sampling frequencies varying by data criticality. Key functions include alarm processing, which generates alerts based on predefined limits, and initial validation through range checks, rate-of-change assessments, and communication error detection to ensure data accuracy and integrity~\cite{cavalari2024enhanced}.

Prior to delving into the detailed analysis of individual attack vectors, it is essential to clarify that GenAI assumes a comprehensive, integrated role within the architecture of the framework. This functionality stands in contrast to the simplistic notion of GenAI operating as a tool with a singular purpose. First, it operates as a language model that interprets natural language descriptions of system states, operational rules, and component relationships, enabling operators and engineers to interact with the detection system using familiar terminology rather than requiring specialized query languages. Second, it functions as a multimodal reasoning engine that synthesizes information from diverse sources including numerical measurements from state estimators, visual representations in HMI displays, and textual operational logs, to identify inconsistencies that span multiple data modalities. Third, it acts as an anomaly classifier that evaluates whether observed conditions violate physical laws, operational constraints, or expected behavioral patterns, subsequently categorizing detected anomalies into specific threat classes such as stealth attacks, database manipulations, or display corruptions. These roles are not implemented as separate modules but rather emerge inherently from GenAI's capabilities when properly guided through structured prompts and domain-specific knowledge. This integrated approach distinguishes the suggested framework from conventional AD systems that typically address either numerical analysis or visual inspection, but rarely within a unified reasoning process.

The following step involves the application of weighted least-squares (WLS) SE, which constitutes the mathematical foundation of EMSs. This technique converts raw measurements into a coherent and dependable depiction of system voltage magnitudes and angles. To maintain data quality, the BDD is conducted employing a statistical $\chi^2$ test. The objective function of the WLS, presumed to adhere to the $\chi^2$ distribution, is evaluated against a predefined threshold. Exceeding the threshold indicates significant errors. An identification of faulty measurements is then carried out through normalized residual analysis, typically involving the exclusion of the data point with the highest residual value and recalculating the state until the function remains below the threshold. Such a methodology is imperative due to the presence of noise or malicious data corruption. Concurrently, the topology processing module examines the statuses of CBs and switches to develop an accurate representation of the network model. It verifies the actual configuration of energized components, which is crucial for large-scale systems where switching is a frequent occurrence. Finally, measurement filtering enhances the quality of the incoming data by reducing noise while maintaining system responsiveness, ensuring that only validated measurements contribute to subsequent monitoring or control algorithms~\cite{pliatsios2020survey}. Then, this data is stored in the EMS Database that serves as a repository for validated electrical network information, including bus voltages, flows of power, and generator outputs. It reflects the best estimate of actual system conditions, using the SE and BDD. Then, the results from analytical applications such as contingency analysis for system security, optimal PF solutions for economic efficiency, and historical trending for long-term analysis and regulatory compliance are stored in the RTDB that acts as the high-performance data hub for all operational applications, maintaining the current state of the system with minimal latency~\cite{choi2024generative, choi2024egridgpt}.
It stores application results that provide the validated system state, optimal PF calculations that determine electrical quantities throughout the network, and contingency analysis results that assess system security~\cite{lin2023synchrophasor}. Each data point comprises quality flags, timestamps, and source information, thereby facilitating the appropriate use by subsequent processes. The RTDB is required to handle rapid updates from various sources while delivering consistent data views to a variety of client applications, necessitating advanced synchronization control and data consistency mechanisms. Finally, the HMI display  portion provides the critical link between the automated systems and human operators who fundamentally remain responsible for system operation. Operator screens present graphical representations of the power system through various visualization modes including single line diagrams (SLDs) that show the electrical connectivity and current state, geographic displays that map the physical location of equipment and current conditions, and trending charts that reveal temporal patterns and help operators forecast future conditions. The alarm presentation system must help operators quickly identify the root cause among possibly hundreds of cascading alerts. Control action interfaces enable operators to issue commands such as opening or closing CBs, adjusting generator setpoints to modify power output, or changing control parameters to influence automatic control behavior. These interfaces include protective mechanisms and confirmation dialogues to prevent accidental actions that could jeopardize system stability.

The modern EMS faces sophisticated cyber threats that exploit vulnerabilities at different stages of the data processing pipeline. Understanding these attack vectors is crucial for developing comprehensive defense mechanisms that protect the integrity of power system operations~\cite{choi2024egridgpt}. To recap, Table~\ref{tab:attack_points_solutions} demonstrates these attack vectors with their proposed AD solutions. 
\begin{table}[!h]
\centering
\caption{Different attack points and their proposed ADSs.}
\label{tab:attack_points_solutions}
\begin{tabular}{|c|c|c|}
\hline
\textbf{Attack Point \#} & \textbf{Description} & \textbf{Proposed Solution} \\
\hline
1 & \makecell{Stealth attacks} & \makecell{GenAI-based AD} \\
2 & \makecell{FDI attacks} & \makecell{GenAI-based AD} \\
3 & \makecell{HMI screen corruption\\ attacks/errors} & \makecell{SoM-GI-based AD} \\
\hline
\end{tabular}
\end{table}
The next section shows these attack models as well as the proposed AD solutions on the GenAI concept and their results and discussion based on the implementations in GenAI tools in detail. Further, the test system is an IEEE 14-bus system which is considered for all attack points and scenarios included in different steps.

\subsection{GenAI Implementation Methodology and Prompt Engineering}
\label{subsec:genai_implementation}
This subsection provides a detailed explanation of how GenAI technology is implemented within the AD framework, addressing the fundamental question of whether LLMs are fine-tuned on power system data or employed through structured prompting methodologies.

\subsubsection{Model Selection and Deployment Strategy}
This proposed framework utilizes Anthropic Claude Pro~\cite{anthropic}, a state-of-the-art LLM, in its pre-trained configuration without domain-specific fine-tuning. This design choice reflects several strategic considerations that distinguish the approach from traditional ML methodologies commonly applied to power system security. Rather than fine-tuning the model on extensive EMS-specific datasets, an approach that would require collecting, labeling, and curating thousands of operational scenarios and attack patterns, the model's existing knowledge leverages a base that already encompasses fundamental physics principles, electrical engineering concepts, mathematical reasoning capabilities, and visual semantic understanding. This pre-trained knowledge provides a robust foundation for understanding power system behavior without the time-intensive and resource-demanding process of model fine-tuning. The decision to avoid fine-tuning offers several practical advantages for EMS applications. First, it significantly reduces the deployment timeline from months to weeks, as organizations do not need to collect extensive training datasets before implementation. Second, it maintains the model's general reasoning capabilities and broad knowledge base, preventing the overfitting issues that can occur when models are narrowly trained on specific datasets. Third, it enables rapid updates to operational rules and constraints by simply modifying prompts rather than retraining models, a critical advantage for adapting to evolving grid configurations and emerging threat landscapes.

\subsubsection{Structured Prompt Engineering Framework}
The core of this methodology lies in systematic prompt engineering that transforms the general-purpose language model into a specialized EMS AD system. These prompts are carefully structured to provide the model with four complementary types of information that collectively enable sophisticated reasoning about power system states and potential anomalies including numerical data integration (i.e., SE, power flow results), power system rules and constraints (e.g., Kirchhoff's Current Law (KCL)), component descriptions for HMI analysis (e.g., red square markers indicate closed CBs at transmission line terminals), and reference scenarios and in-context learning (ICL) which can be generated using the contingency analysis based on the power flow processes in PowerWorld Simulator. These reference cases serve as benchmark scenarios that guide the model's interpretation of subsequent test scenarios, effectively creating a learned context without modifying the underlying model parameters. Hence, the prompt-based methodology offers several distinct advantages compared to fine-tuning LLMs on power system-specific datasets, including rapid deployment, transparency and interpretability, easy adaptation, preservation of general knowledge, and reduced data requirements.

To clarify the implementation methodology, this study employs Anthropic Claude Pro as the primary GenAI tool, as referenced in~\cite{anthropic}. The simulation process consists of several key steps. Initially, the Claude Pro model was trained using PF results, SE rules, and normal operational scenarios from the IEEE 14-bus system. Then, Excel files containing both normal and attacked data scenarios are given as inputs. Further, domain-specific prompts incorporating power system knowledge and detection rules were developed. The analysis was executed where the trained model examines uploaded data to detect anomalies based on learned patterns and physical constraints. Finally, the framework validated against multiple attack scenarios to verify detection performance. Throughout this paper, the ``Implementation in GenAI Tool'' subsections serve multiple critical purposes. They provide transparency by demonstrating actual interactions between researchers and the AI system with exact prompts and responses and enable reproducibility by offering sufficient methodological detail, then validate the approach by showing real GenAI outputs rather than theoretical capabilities. Hence, it can bridge the gap between theoretical framework and practical implementation, making the methodology accessible to other researchers in this domain.

\section{Attack Vectors and Experimental Validation} \label{implementation-section}
\subsection{Attack Point \#1: Stealth Attacks (Gray Dashed Line)}
Stealth attacks represent a particularly malicious threat to the security of EMSs, as they are meticulously engineered to bypass conventional BDD mechanisms on which operators depend to preserve the integrity of the system. These attacks are characterized by their ability to remain undetected by standard monitoring protocols, thus presenting a significant challenge to the protective measures employed within EMS frameworks. By leveraging vulnerabilities in AD methodologies, they weaken the robustness of systems to preserve operational stability and security~\cite{xu2024globally}. These attacks utilize the mathematical foundations of SE algorithms by injecting carefully crafted false measurements that maintain consistency with the power system's physical laws and network topology. When executed successfully, a stealth attack manipulates the estimated system state while satisfying all residual tests and WLS criteria, making the corrupted data appear legitimate to conventional BDD systems. The complexity of these attacks lies in their ability to manipulate critical operational parameters (e.g., bus voltages and  PFs), without triggering alarms, potentially leading operators to make incorrect decisions based on falsified system conditions. This misleading essence makes stealth attacks particularly dangerous, as they can persist undetected for extended periods while progressively degrading system reliability or creating opportunities for more severe disruptions. Also, these stealth attacks can make major gradual impacts on power systems including economic dispatch inefficiencies, deterioration of operator confidence, compromised grid resilience, degraded system reliability over time, cascading failure and blackouts, and compromised system simulations and estimates~\cite{ashrafuzzaman2020detecting, guo2023residual, zhou2025worst}. 

\begin{figure}[!h]
\centerline{\includegraphics[width=1.0\columnwidth]{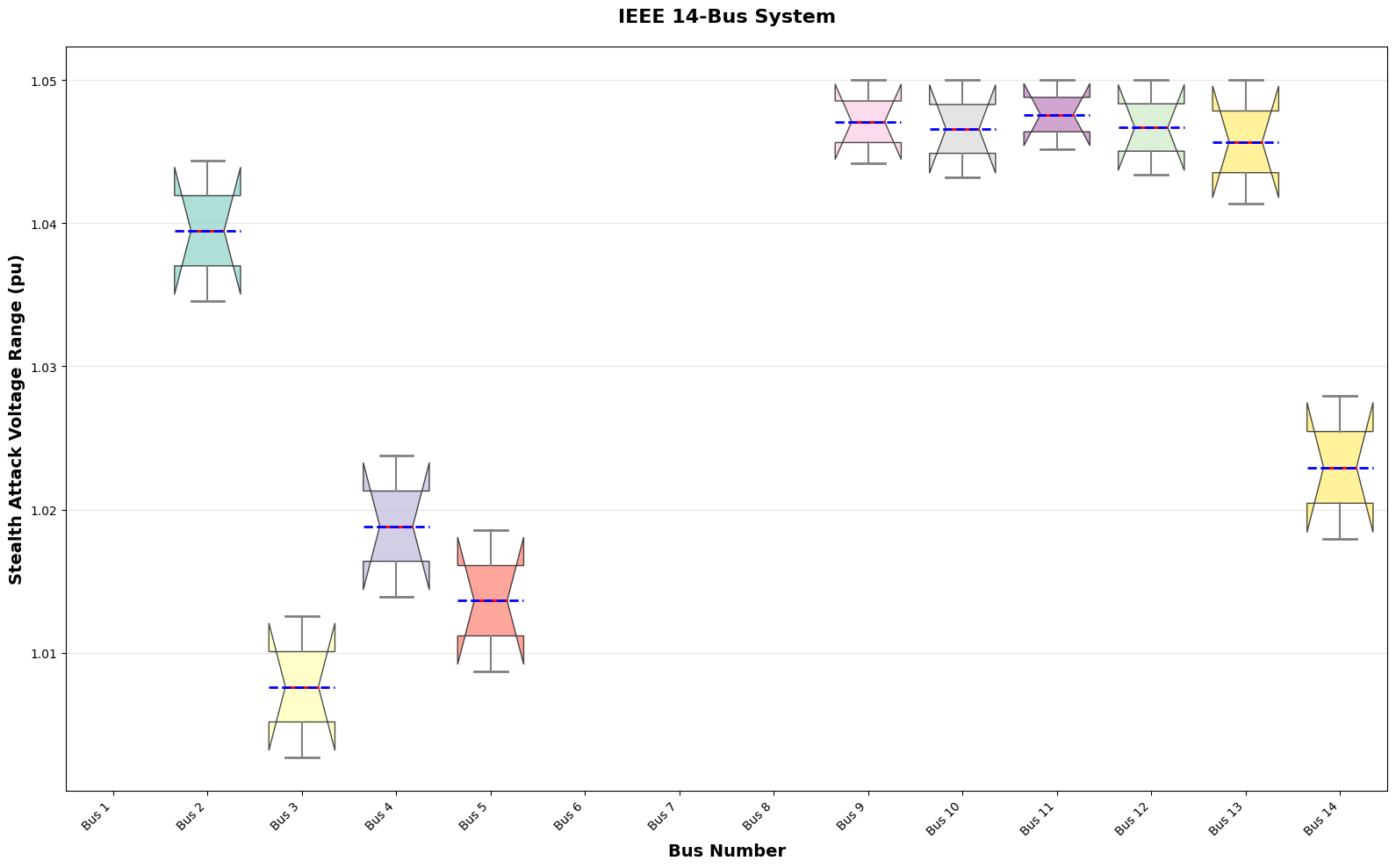}}
\caption{\centering The ranges of stealth attacks for different buses according to the bus voltage magnitudes for an IEEE 14-bus system.}
\label{stealth_ieee_14bus_0.95_1.05}
\end{figure}

\paragraph{A Mathematical Construction of SE and Stealth Attacks}
Following the collection of PF measurements, power input data, and voltage magnitude information gathered from the system's buses by SCADA units, the initiation of the static SE process takes place. The SE algorithm aims to determine the state vector $\mathbf{x} \in \mathbb{R}^n$, which includes both phase angles and voltage magnitudes across different buses, where $n = 2k - 1$ with $k$ representing the total number of buses. In the context of AC static SE, the relationship between the state vector $\mathbf{x}$ and measurements follows the nonlinear model as Eq.~(\ref{SE_AC_nonlinear})~\cite{ashrafuzzaman2020detecting}:
\begin{equation} \label{SE_AC_nonlinear}
\mathbf{z} = \mathbf{H}\mathbf{x} + \mathbf{e}
\end{equation}
Here, the measurement vector $\mathbf{z} \in \mathbb{R}^m$ comprises readings gathered by SCADA units, with $m$ denoting the quantity of measurements. The nonlinear mapping function $\mathbf{H}(\cdot)$ is derived from the grid's topological structure and characteristics of transmission lines, transformers, and related grid components. The error term $\mathbf{e} \in \mathbb{R}^m$ follows a Gaussian distribution characterized by the covariance matrix $R$. To estimate the state vector $\mathbf{x}$, an iterative WLS algorithm is employed as Eq.~(\ref{SE_AC_State_Vector}):
\begin{equation} \label{SE_AC_State_Vector}
\hat{\mathbf{x}}_k = \hat{\mathbf{x}}_{k-1} + \mathbf{H}_k^{\dagger}(z_k - \mathbf{H}(\mathbf{x}_{k-1}))
\end{equation}
where $\mathbf{H}_k^{\dagger} = (\mathbf{H}_k^T R^{-1} \mathbf{H}_k)^{-1} \mathbf{H}_k^T R^{-1}$ and $\mathbf{H}_k$ represent the Jacobian matrix of $\mathbf{H}$ evaluated at iteration $k$. Under Gaussian noise assumptions, this WLS approach yields optimal results. Upon convergence, achieved when $\|\hat{\mathbf{x}}_k - \hat{\mathbf{x}}_{k-1}\| < \delta$ for a predefined small threshold $\delta > 0$, the analysis of resulting residuals detects potential measurement anomalies by verifying Gaussian properties. Such anomalies in data might stem from natural failures (e.g., sensor malfunctions or communication disruptions, or potentially from deliberate FDI attacks). Standard detection methods typically employ $\chi^2$ testing for identifying anomalous data. Additionally, a DC SE variant exists where only phase angles require estimation, with voltages assumed to be at unity (1 p.u.). This simplified model neglects line resistances and assumes small phase angle differences between buses, resulting in a linear regression framework as Eq.~(\ref{SE_DC_vector})~\cite{zhang2023limitation}:
\begin{equation} \label{SE_DC_vector}
\mathbf{z} = \mathbf{H}\mathbf{x} + \mathbf{e}
\end{equation}
where $\mathbf{z} \in \mathbb{R}^m$ represents the measurement vector, $\mathbf{H} \in \mathbb{R}^{m \times n}$ is the measurement Jacobian matrix, $\mathbf{x} \in \mathbb{R}^n$ is the state vector, and $\mathbf{e} \sim \mathcal{N}(\mathbf{0}, \mathbf{R})$ is the measurement error vector with covariance matrix $\mathbf{R}$. The SE, $\hat{\mathbf{x}}$, is typically obtained through the WLS through Eq.~(\ref{eq:state_estimate_2}):
\begin{equation} \label{eq:state_estimate_2}
\hat{\mathbf{x}} = (\mathbf{H}^T \mathbf{R}^{-1} \mathbf{H})^{-1} \mathbf{H}^T \mathbf{R}^{-1} \mathbf{z}
\end{equation}
After SE, the measurement residual vector $\mathbf{r}$ is calculated as Eq.~(\ref{eq:residual_1}):
\begin{equation} \label{eq:residual_1}
\mathbf{r} = \mathbf{z} - \mathbf{H}\hat{\mathbf{x}}
\end{equation}
Traditional BDD mechanisms typically use the $\chi^2$-test on the residual as shown in Eq.~(\ref{eq:chi_squared_1}):
\begin{equation} \label{eq:chi_squared_1}
J(\mathbf{x}) = \mathbf{r}^T \mathbf{R}^{-1} \mathbf{r} \leq \tau
\end{equation}
where $\tau$ is a threshold value derived from the $\chi^2$ distribution with appropriate degrees of freedom. A stealth attack involves the addition of an attack vector $\mathbf{a}$ to the measurement vector which is given in Eq.~(\ref{eq:attacked_measurements_1}):
\begin{equation} \label{eq:attacked_measurements_1}
\mathbf{z}_a = \mathbf{z} + \mathbf{a}
\end{equation}
The essential concept for stealth attacks is to construct $\mathbf{a}$ in the column space of $\mathbf{H}$ (Eq.~(\ref{eq:attack_vector_1})):
\begin{equation} \label{eq:attack_vector_1}
\mathbf{a} = \mathbf{H}\mathbf{c}
\end{equation}
where $\mathbf{c}$ is an arbitrary vector in the state space. When such an attack is applied, the new state estimate becomes as follows:
\begin{equation} \label{eq:attacked_state_1}
\hat{\mathbf{x}}_a = \hat{\mathbf{x}} + \mathbf{c}
\end{equation}
Which the residual crucially remains unchanged as shown in Eq.~(\ref{eq:unchanged_residual_1}):
\begin{equation} \label{eq:unchanged_residual_1}
\mathbf{r}_a = \mathbf{z}_a - \mathbf{H}\hat{\mathbf{x}}_a = \mathbf{z} + \mathbf{H}\mathbf{c} - \mathbf{H}(\hat{\mathbf{x}} + \mathbf{c}) = \mathbf{z} - \mathbf{H}\hat{\mathbf{x}} = \mathbf{r}
\end{equation}
This characteristic of the mathematical model enables the attack approach to evade conventional mechanisms designed for detecting erroneous data, as $J(\mathbf{x}_a) = J(\mathbf{x}) \leq \tau$~\cite{ashrafuzzaman2020detecting}. According to a case study for an IEEE 14-bus system, 300 attack points are applied to the system at each bus to find the range of stealth attacks for different buses, forced by the NERC regulation in terms of bus voltage magnitude violation, which states that the bus voltages should fall within the range of 0.95--1.05 p.u. to meet the requirements~\cite{NERC2020ReactivePower}. According to this process, a sample of the stealth attack range for Bus 2 is illustrated in Table~\ref{stealth_bus2_range}. All these attack points are also injected to all other buses to find the range of stealth attack as illustrated in Fig.~\ref{stealth_ieee_14bus_0.95_1.05}.
\begin{table}[!h]
\centering
\scriptsize
\caption{A part of attack points applied to Bus 2 to find the stealth attack range.}
\label{stealth_bus2_range}
\begin{tabular}{|c|c|c|c|c|c|c|l|}
\hline
\textbf{Bus} & \textbf{Attack\_Vm} & \textbf{Original\_Vm} & \textbf{Detected} & \textbf{Anomaly Detection} \\
\hline
2 & 1.033277592 & 1.044446943 & TRUE & Bad data detected \\
\hline
2 & 1.033779264 & 1.044446943 & TRUE & Bad data detected \\
\hline
2 & 1.034280936 & 1.044446943 & TRUE & Bad data detected \\
\hline
2 & 1.034782609 & 1.044446943 & \textcolor{red}{FALSE} & Stealth attack \\
\hline
2 & 1.035284281 & 1.044446943 & \textcolor{red}{FALSE} & Stealth attack \\
\hline
2 & 1.035785953 & 1.044446943 & \textcolor{red}{FALSE} &  Stealth attack \\
\hline
2 & 1.036287625 & 1.044446943 & \textcolor{red}{FALSE} &  Stealth attack \\
\hline
2 & 1.036789298 & 1.044446943 & \textcolor{red}{FALSE} &  Stealth attack \\
\hline
2 & 1.03729097 & 1.044446943 & \textcolor{red}{FALSE} &  Stealth attack \\
\hline
2 & 1.037792642 & 1.044446943 & \textcolor{red}{FALSE} &  Stealth attack \\
\hline
2 & 1.038294314 & 1.044446943 & \textcolor{red}{FALSE} &  Stealth attack \\
\hline
2 & 1.038795987 & 1.044446943 & \textcolor{red}{FALSE} &  Stealth attack \\
\hline
2 & 1.039297659 & 1.044446943 & \textcolor{red}{FALSE} &  Stealth attack \\
\hline
2 & 1.039799331 & 1.044446943 & \textcolor{red}{FALSE} &  Stealth attack \\
\hline
2 & 1.040301003 & 1.044446943 & \textcolor{red}{FALSE} &  Stealth attack \\
\hline
2 & 1.040802676 & 1.044446943 & \textcolor{red}{FALSE} &  Stealth attack \\
\hline
2 & 1.041304348 & 1.044446943 & \textcolor{red}{FALSE} &  Stealth attack \\
\hline
2 & 1.04180602 & 1.044446943 & \textcolor{red}{FALSE} &  Stealth attack \\
\hline
2 & 1.042307692 & 1.044446943 & \textcolor{red}{FALSE} &  Stealth attack \\
\hline
2 & 1.042809365 & 1.044446943 & \textcolor{red}{FALSE} &  Stealth attack \\
\hline
2 & 1.043311037 & 1.044446943 & \textcolor{red}{FALSE} &  Stealth attack \\
\hline
2 & 1.043812709 & 1.044446943 & \textcolor{red}{FALSE} &  Stealth attack \\
\hline
2 & 1.044314381 & 1.044446943 & \textcolor{red}{FALSE} &  Stealth attack \\
\hline
2 & 1.044816054 & 1.044446943 & TRUE & Bad data detected \\
\hline
2 & 1.045317726 & 1.044446943 & TRUE & Bad data detected \\
\hline
2 & 1.045819398 & 1.044446943 & TRUE & Bad data detected \\
\hline
\end{tabular}
\end{table}
The presented visualization reveals the feasible ranges across the IEEE 14-bus system that successfully evade the $\chi^2$ as a BDD system while maintaining system observability. Further, a numerical representation of stealth attack ranges including the start point, end point, width of the range, and original bus voltages based on the numerical representation is given in Table~\ref{stealth_arrack_range_numerical}. 
\begin{table*}[!h]
\centering
\caption{\centering Stealth attack ranges results for an IEEE 14-bus system based on start/end points of attacks along with the original voltage before applying the NERC regulation.}
\label{stealth_arrack_range_numerical}
\begin{tabular}{|c|c|c|c|c|c|}
\hline
\makecell{\textbf{Bus No.}} & \makecell{\textbf{Bus type}} & \makecell{\textbf{Stealth attack}\\\textbf{\_start point}} & \makecell{\textbf{Stealth attack}\\\textbf{\_end point}} & \makecell{\textbf{Stealth attack}\\\textbf{\_width}} & \makecell{\textbf{Original}\\ \textbf{voltage}} \\
\hline
1 & Slack & N/A & N/A & N/A & 1.061987 \\
\hline
2 & Generator & 1.034569138 & 1.044388778 & 0.009819639 & 1.044446943 \\
\hline
3 & Generator & 1.002705411 & 1.01252505 & 0.009819639 & 1.012590754 \\
\hline
4 & Load & 1.013927856 & 1.023747495 & 0.009819639 & 1.023762973 \\
\hline
5 & Load & 1.008717435 & 1.018537074 & 0.009819639 & 1.018577246 \\
\hline
6 & Generator & N/A & N/A & N/A & 1.069063452 \\
\hline
7 & Load & N/A & N/A & N/A & 1.067836384 \\
\hline
8 & Generator & N/A & N/A & N/A & 1.093069739 \\
\hline
9 & Load & 1.044188377 & 1.05 & 0.005811623 & 1.054053823 \\
\hline
10 & Load & 1.043186373 & 1.05 & 0.006813627 & 1.053154865 \\
\hline
11 & Load & 1.045190381 & 1.05 & 0.004809619 & 1.055052848 \\
\hline
12 & Load & 1.043386774 & 1.05 & 0.006613226 & 1.053325644 \\
\hline
13 & Load & 1.041382766 & 1.05 & 0.008617234 & 1.051349563 \\
\hline
14 & Load & 1.017935872 & 1.027955912 & 0.01002004 & 1.027876825 \\
\hline
\end{tabular}
\end{table*}
According to these findings, Bus 3 exhibits the most extensive stealth range, spanning approximately 0.004 to 0.012 p.u. around its nominal value, indicating its elevated vulnerability to stealth attacks due to its network position and measurement redundancy characteristics. Buses 2 and 14 demonstrate similarly broad attack boundaries, with voltage deviations permissible within $\pm$ 0.025 p.u. from their baseline values while remaining undetected. In contrast, bus 5 displays notably constrained attack ranges, suggesting its measurements are more tightly coupled to the system's observable state through the measurement Jacobian matrix. Buses 6, 7, and 8 show no stealth attack ranges within the defined voltage magnitude as well as no range for the slack bus (i.e., Bus 1). While satisfying the stealth constraint $\chi^2 < 89.5$, these diverse exploitable boundaries across the network topology demonstrate that successful stealth attacks must account for bus-specific constraints, measurement configurations, and their contributions to the overall WLS residual, finally exposing the diverse susceptibility profile of the power system to sophisticated network breaches. Now, according to the given information, two scenarios are presented to make a comparison of the traditional BDDs and the GenAI-based ADS according to the stealth attacks. The definition of these attacks is described , then the implementation of GenAI is explained to show the results of the detection process.
\subsubsection{Scenario \#1A: 5-Point Distributed Stealth Attack}
\begin{figure}[!t]
    \centering
    \begin{subfigure}[b]{0.5\textwidth}
        \includegraphics[width=\textwidth]{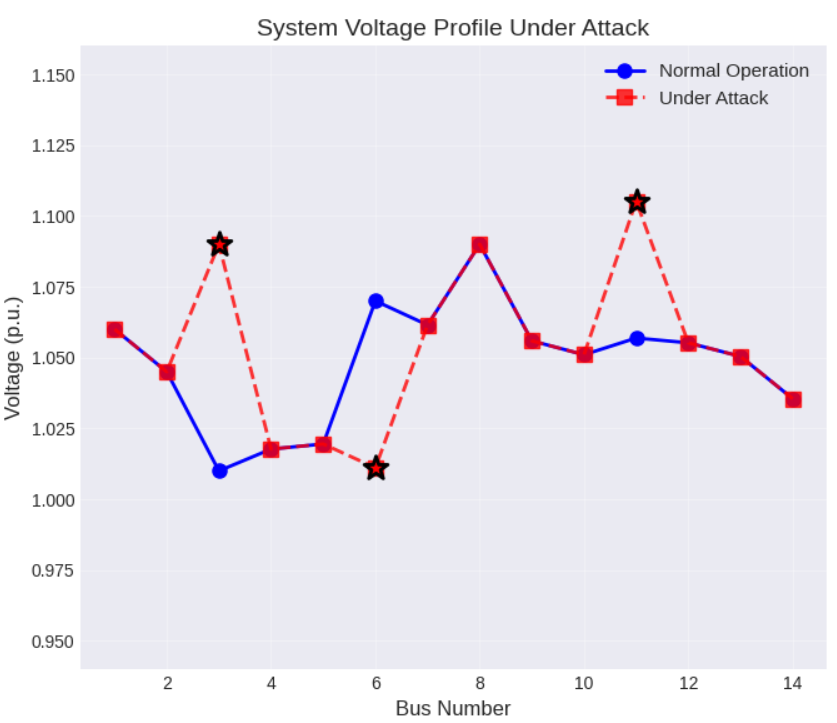}
        \caption{}
        \label{subfig_stealth_1A_voltage}
    \end{subfigure}
    \hfill
    \begin{subfigure}[b]{0.5\textwidth}
        \includegraphics[width=\textwidth]{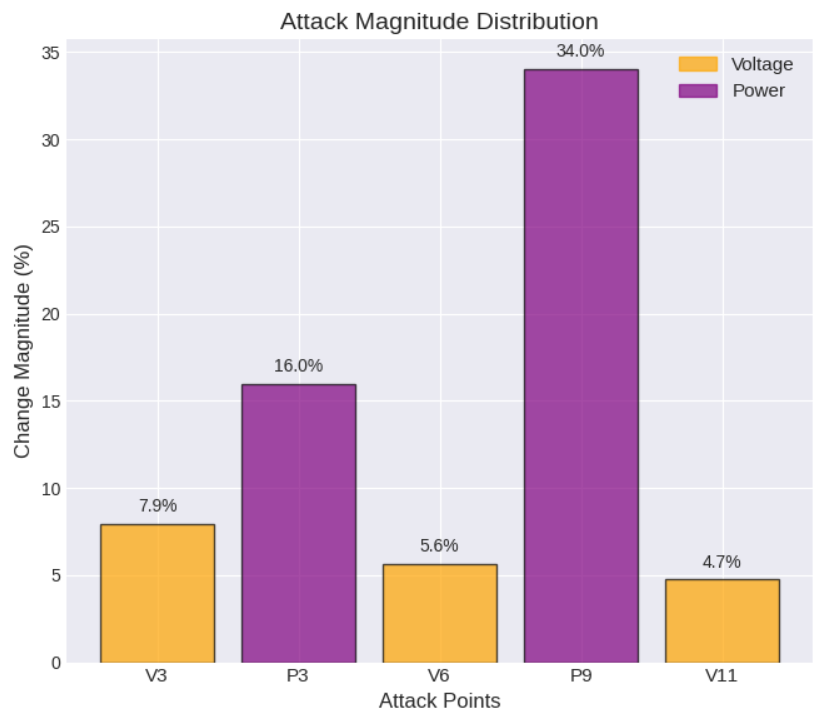}
        \caption{}
        \label{subfig_stealth_1A_percentage}
    \end{subfigure}
    \caption{A visualization of the attack vector in Scenario \#1A, (a) system voltage profile under attack (b) attack magnitude distribution (\%).}
    \label{fig_stealth_1A_attack_vector}
\end{figure}

This scenario represents a 5-point distributed stealth attack (i.e. Eq.~(\ref{scenario1A_attack_vector})) designed to evade the $\chi^2$ test based on the BDD while maintaining all PF constraints. The attack strategically targets five measurement points across the IEEE 14-bus system with coordinated changes as given in the attack vector.
\begin{equation} \label{scenario1A_attack_vector}
\mathbf{a} = \begin{bmatrix}
\Delta V_3 \\ \Delta P_3 \\ \Delta V_6 \\ \Delta P_9 \\ \Delta V_{11}
\end{bmatrix} = \begin{bmatrix}
+0.08 \\ +0.15 \\ -0.06 \\ +0.10 \\ +0.05
\end{bmatrix} \text{ p.u.}
\end{equation}
Further, a visual representation of the attack vector for voltage and active power changes is demonstrated in Fig.~\ref{fig_stealth_1A_attack_vector}. 
As shown, the voltage at bus 3 increases by 7.9\% (from 1.0100 to 1.0900 p.u.), active power at bus 3 increases by 16.0\% (from 0.9399 to 1.0899 p.u.), voltage at bus 6 decreases by 5.6\% (from 1.0711 to 1.0111 p.u.), active power at bus 9 increases by 34.0\% (from 0.2937 to 0.3936 p.u.), and voltage at bus 11 increases by 4.7\% (from 1.0552 to 1.1052 p.u.) with a base active power of 100 MW. To maintain the stealth property and satisfy power balance constraints, the total power injection increase of 0.25 p.u. (from $P_3$ and $P_9$ changes) is strategically distributed as compensation across seven non-attacked buses (buses 2, 4, 5, 10, 12, 13, and 14), with each receiving a reduction of 0.0357 p.u. This distributed compensation mechanism, combined with the opposing voltage changes ($V_6$ decreasing while $V_3$ and $V_{11}$ increase), ensures the attack's $\chi^2$ statistic remains below the detection threshold of 89.5.

\subsubsection{Scenario \#1B: 8-Point Massive Coordinated Attack}
This scenario demonstrates an even more complex 8-point massive coordinated attack (i.e. Eq.~(\ref{scenario1B_attack_vector})) that exploits the fundamental limitations of statistical BDD through measurement manipulation. 
\begin{equation} \label{scenario1B_attack_vector}
\mathbf{a} = \begin{bmatrix}
\Delta V_2 \\ \Delta P_2 \\ \Delta V_4 \\ \Delta P_4 \\ \Delta V_6 \\ \Delta P_9 \\ \Delta V_{11} \\ \Delta P_{13}
\end{bmatrix} = \begin{bmatrix}
+0.09 \\ +0.15 \\ -0.07 \\ -0.13 \\ +0.08 \\ +0.12 \\ -0.06 \\ -0.10
\end{bmatrix} \text{ p.u.}
\end{equation}
This attack simultaneously modifies eight critical measurements across the IEEE 14-bus system as represented in the attack vector. 

\begin{figure}[!h]
    \centering
    \begin{subfigure}[b]{0.5\textwidth}
        \includegraphics[width=\textwidth]{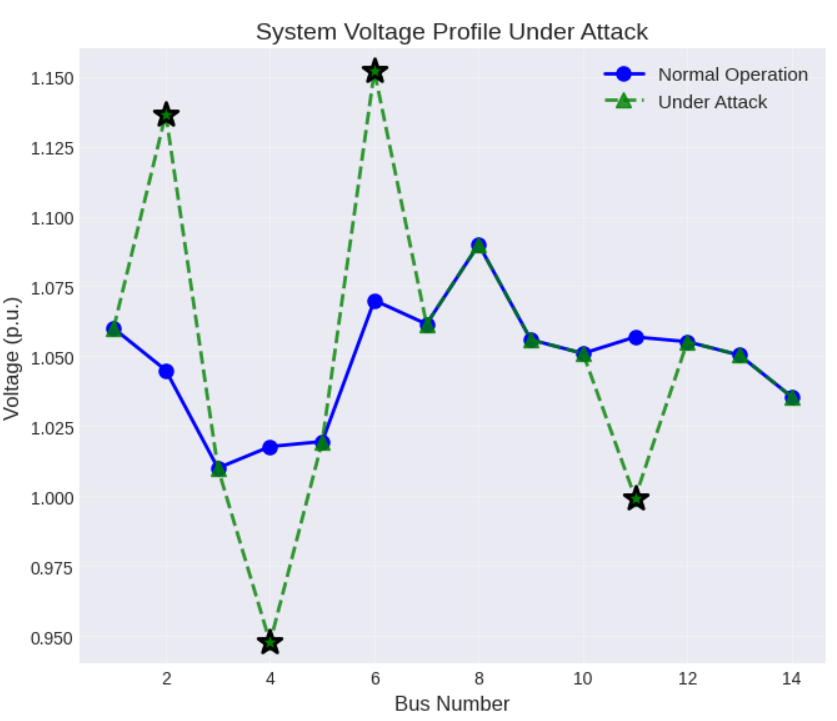}
        \caption{}
        \label{subfig_stealth_1B_voltage}
    \end{subfigure}
    \hfill
    \begin{subfigure}[b]{0.5\textwidth}
        \includegraphics[width=\textwidth]{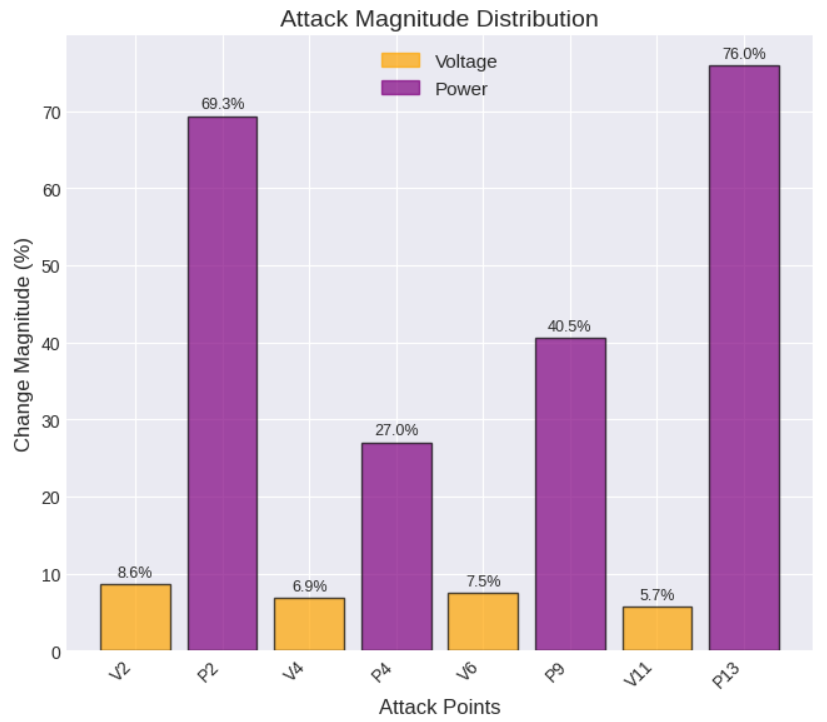}
        \caption{}
        \label{subfig_stealth_1B_percentage}
    \end{subfigure}
    \caption{A visualization of the attack vector in Scenario \#1B, (a) system voltage profile under attack (b) attack magnitude distribution (\%).}
    \label{fig_stealth_1B_attack_vector}
\end{figure}

Also, Fig.~\ref{fig_stealth_1B_attack_vector} illustrates a visual representation of the attack vector for this scenario according to these concurrent attacks at 8 points with changes in voltage magnitudes and active power values. 
As demonstrated based on the similar attack description with Scenario \#1A, the voltage at bus 2 increases by 8.6\% (from 1.0466 to 1.1366 p.u.) with its active power increasing by 69.3\% (from 0.2163 to 0.3663 p.u.); voltage at bus 4 decreases by 6.9\% (from 1.0176 to 0.9476 p.u.) with power decreasing by 27.0\% (from 0.4809 to 0.3509 p.u.); voltage at bus 6 increases by 7.5\% (from 1.0719 to 1.1519 p.u.); active power at bus 9 increases by 40.5\% (from 0.2960 to 0.4160 p.u.); voltage at bus 11 decreases by 5.7\% (from 1.0594 to 0.9994 p.u.); and active power at bus 13 experiences a dramatic 76.0\% decrease (from 0.1316 to 0.0316 p.u.). The attack's sophistication lies in its perfect coordination of opposing changes---positive power changes (+0.15 at $P_2$, +0.12 at $P_9$) are nearly balanced by negative changes ($-0.13$ at $P_4$, $-0.10$ at $P_{13}$), resulting in a minimal net power change of only 0.04 p.u. Additionally, random noise is injected at non-attacked buses (1, 5, 7, 8, 10, 12, and 14) to further conceal the attack pattern, achieving a $\chi^2$ test statistic of 67.3, still comfortably below the 89.5 detection threshold while creating a highly abnormal but mathematically valid system state.
\vspace{2mm}

\textbf{Implementation in GenAI Tool} 

\begin{responsebox}
\small
The state estimation problem determines the system state vector $\mathbf{x} = [V_1, \theta_1, \ldots, V_{14}, \theta_{14}]^T$ from measurements $\mathbf{z}$ using Weighted Least Squares (WLS):
\begin{equation*}
\text{Minimize: } J(\mathbf{x}) = [\mathbf{z} - \mathbf{h}(\mathbf{x})]^T \mathbf{R}^{-1} [\mathbf{z} - \mathbf{h}(\mathbf{x})]
\end{equation*}
Where:
\begin{itemize}
\item $\mathbf{z} \in \mathbb{R}^m$: Measurement vector
\item $\mathbf{h}(\mathbf{x})$: Non-linear measurement functions from power flow equations
\item $\mathbf{R}$: Measurement error covariance matrix
\end{itemize}

The bad data detection test evaluates:
\begin{equation*}
\chi^2 = \mathbf{r}^T \mathbf{R}^{-1} \mathbf{r}
\end{equation*}
where $\mathbf{r} = \mathbf{z} - \mathbf{h}(\hat{\mathbf{x}})$ is the measurement residual vector. The $\chi^2$ threshold is calculated based on the statistical test framework for bad data detection. The threshold can be derived as follows:

\noindent The detection system operates on an enhanced feature space consisting of:
\begin{equation*}
\mathbf{F} = [\mathbf{z}_{direct}, \mathbf{z}_{derived}] \in \mathbb{R}^{71}
\end{equation*}
where the 71-dimensional feature vector comprises:
\begin{itemize}
\item Direct measurements: 42 dimensions
    \begin{itemize}
    \item Voltage measurements: $V_1, V_2, \ldots, V_{14}$ (14 dimensions)
    \item Active power measurements: $P_1, P_2, \ldots, P_{14}$ (14 dimensions)
    \item Reactive power measurements: $Q_1, Q_2, \ldots, Q_{14}$ (14 dimensions)
    \end{itemize}
\item Statistical features: 8 dimensions
    \begin{itemize}
    \item $[\mu_V, \sigma_V, \min(V), \max(V), \mu_P, \sigma_P, \min(P), \max(P)]$
    \end{itemize}
\item Correlation features: 3 dimensions
    \begin{itemize}
    \item $[\rho_{V,P}, \rho_{V,Q}, \rho_{P,Q}]$
    \end{itemize}
\item Physics-based features: 5 dimensions
    \begin{itemize}
    \item $[P_{total}, Q_{total}, PF, V_{stability}, P_{imbalance}]$
    \end{itemize}
\item Gradient features: 13 dimensions
    \begin{itemize}
    \item $\nabla V = [V_2-V_1, V_3-V_2, \ldots, V_{14}-V_{13}]$
    \end{itemize}
\item Second-order features: Additional dimensions
\end{itemize}

Total: $42 + 8 + 3 + 5 + 13 = 71$ dimensions

For the enhanced feature space, the test statistic follows a chi-square distribution:
\begin{equation*}
\chi^2 = (\mathbf{F} - \boldsymbol{\mu}_F)^T \boldsymbol{\Sigma}_F^{-1} (\mathbf{F} - \boldsymbol{\mu}_F) \sim \chi^2_{71}
\end{equation*}
where:
\begin{itemize}
\item $\boldsymbol{\mu}_F \in \mathbb{R}^{71}$: Mean of the feature vector under normal operation
\item $\boldsymbol{\Sigma}_F \in \mathbb{R}^{71 \times 71}$: Covariance matrix of features
\item Degrees of freedom: $df = 71$
\end{itemize}

\noindent For a significance level $\alpha = 0.05$ (95\% confidence level), the threshold is:
\begin{equation*}
\chi^2_{threshold} = \chi^2_{1-\alpha, df} = \chi^2_{0.95, 71}
\end{equation*}
\noindent Using the inverse chi-square cumulative distribution function:
\begin{equation*}
\chi^2_{0.95, 71} = F^{-1}_{\chi^2_{71}}(0.95) = 89.5
\end{equation*}
This means:
\begin{equation*}
P(\chi^2 < 89.5 | H_0) = 0.95
\end{equation*}
\noindent where $H_0$ represents the null hypothesis of no bad data (normal operation). Under normal operating conditions:
\begin{itemize}
\item 95\% of measurements will have $\chi^2 < 89.5$
\item Only 5\% will exceed this threshold (false alarm rate)
\item If $\chi^2 > 89.5$, bad data is detected with 95\% confidence
\item If $\chi^2 < 89.5$, measurements are considered statistically consistent $\rightarrow$ \textbf{Normal operation} (no bad data)
\end{itemize}

\paragraph{Scenario \#1A: 5-Point Measurement Changes}
The system experiences the following measurement changes from baseline:
\begin{center}
\begin{tabular}{|l|l|l|}
\toprule
Bus & Measurement & Change \\
\midrule
3 & $V_3$ & +0.0800 p.u. (+7.9\%) \\
3 & $P_3$ & +0.1500 p.u. (+16.0\%) \\
6 & $V_6$ & -0.0600 p.u. (-5.6\%) \\
9 & $P_9$ & +0.1000 p.u. (+34.0\%) \\
11 & $V_{11}$ & +0.0500 p.u. (+4.7\%) \\
\bottomrule
\end{tabular}
\end{center}

Additionally, power measurements at buses $2, 4, 5, 10, 12, \\ 13, 14$ show small decreases of 0.0357 p.u. each.
\vspace{2mm}

\textbf{State Estimation Analysis} Based on the Power Flow Solution:
\begin{itemize}
\item The state estimator successfully converges
\item All power flow equations are satisfied:
\begin{align*}
P_i &= V_i \sum_{j} V_j (G_{ij} \cos(\theta_i - \theta_j) + B_{ij} \sin(\theta_i - \theta_j)) \quad \checkmark \\
Q_i &= V_i \sum_{j} V_j (G_{ij} \sin(\theta_i - \theta_j) - B_{ij} \cos(\theta_i - \theta_j)) \quad \checkmark
\end{align*}
\item Power balance maintained: $\sum \Delta P \approx 0$
\end{itemize}
\vspace{2mm}

\textbf{Chi-Square Test Result}

Statistical Analysis:
\begin{equation*}
\chi^2 = (\mathbf{z} - \mathbf{h}(\hat{\mathbf{x}}))^T \mathbf{R}^{-1} (\mathbf{z} - \mathbf{h}(\hat{\mathbf{x}})) = 42.8
\end{equation*}

Threshold: $\chi^2_{0.05,71} = 89.5$
\vspace{1mm}

\textbf{Chi-Square Conclusion: NORMAL OPERATION} $\checkmark$
\vspace{1mm}

Since $42.8 < 89.5$, the chi-square test indicates:
\begin{itemize}
\item No bad data detected
\item Measurements are statistically consistent
\item Scenario represents valid system operation
\end{itemize}
\vspace{2mm}

\textbf{GenAI Analysis and Detection}
The correlation analysis reveals:
\begin{align*}
\text{Expected V-P correlation at bus 3:} \quad &\rho = 0.3 \pm 0.1 \\
\text{Observed correlation:} \quad &\rho = 0.8 \\
\text{Deviation:} \quad &5 \text{ standard deviations}
\end{align*}

This dramatic correlation shift has never occurred in training samples of normal operation.

GenAI computes power-voltage sensitivities~\cite{yoshizawa2021voltage}:
\begin{equation*}
\frac{\partial P_3}{\partial V_3} = \frac{\Delta P_3}{\Delta V_3} = \frac{0.15}{0.08} = 1.875 \text{ p.u.}
\end{equation*}

Which the historical range is $[0.77, 1.07]$ p.u. This sensitivity exceeds physical limits for the network topology and loading conditions.

According to the entropy analysis, the uniform distribution across 7 buses indicates artificial coordination rather than natural load variation. In terms of the voltage gradient analysis, $\nabla V$ between buses 2-3 is 0.045 p.u. while the maximum historical gradient is 0.020 p.u. This violates learned spatial coherence patterns.
\vspace{2mm}

\noindent \textbf{GenAI Classification: ATTACK DETECTED}
\vspace{2mm}

\noindent \textbf{Reasoning:} Despite satisfying all power flow equations and passing the chi-square test, the measurement pattern exhibits:
\begin{enumerate}
\item Fabricated correlation changes
\item Physically implausible sensitivities
\item Artificial load distribution
\item Coordinated changes at electrically distant buses
\end{enumerate}

\paragraph{Scenario \#1B: 8-Point Measurement Changes}
The system shows the following measurement changes:
\begin{center}
\begin{tabular}{|l|l|l|}
\toprule
Bus & Measurement & Change \\
\midrule
2 & $V_2$ & +0.0900 p.u. (+8.6\%) \\
2 & $P_2$ & +0.1500 p.u. (+69.3\%) \\
4 & $V_4$ & -0.0700 p.u. (-6.9\%) \\
4 & $P_4$ & -0.1300 p.u. (-27.0\%) \\
6 & $V_6$ & +0.0800 p.u. (+7.5\%) \\
9 & $P_9$ & +0.1200 p.u. (+40.5\%) \\
11 & $V_{11}$ & -0.0600 p.u. (-5.7\%) \\
13 & $P_{13}$ & -0.1000 p.u. (-76.0\%) \\
\bottomrule
\end{tabular}
\end{center}
\vspace{2mm}

\textbf{State Estimation Analysis} based on the Power Flow Solution:
\begin{itemize}
\item State estimation converges after 8 iterations
\item All power flow constraints satisfied
\item Net power change: $\sum \Delta P = 0.04$ p.u. (minimal)
\end{itemize}
\vspace{2mm}

\textbf{Chi-Square Test Result}

Statistical Analysis:
\begin{equation*}
\chi^2 = 67.3
\end{equation*}

Threshold: $\chi^2_{0.05,71} = 89.5$
\vspace{2mm}

\textbf{Chi-Square Conclusion: NORMAL OPERATION} $\checkmark$
\vspace{2mm}

Since $67.3 < 89.5$, chi-square indicates:
\begin{itemize}
\item No bad data detected
\item Measurements pass statistical consistency check
\item Scenario classified as legitimate system state
\end{itemize}
\vspace{2mm}

\textbf{GenAI Analysis and Detection}
GenAI's comprehensive analysis reveals severe anomalies:

The generator Ramp Rate Analysis (Bus 2):
\begin{align*}
\text{Power change:} \quad &+69.3\% \text{ instantaneous} \end{align*}

It is recognized that this is physically impossible for any generator.

According to the Load Behavior Analysis (Bus 13), the following completely violates ZIP (i.e., impedance, current, power) load model physics. If voltage remains nearly constant, power should also remain relatively constant.
\begin{align*}
\text{Power reduction:} \quad &76\% \\
\text{Voltage change:} \quad &\text{minimal} \\
\text{Expected from load model:} \quad &P \propto V^{\alpha} \text{ where } \alpha \in [0.5, 2.0]
\end{align*}



According to the attention mechanism results, the attention weights reveal:
\begin{align*}
\text{Attention}(Q,K,V) = \text{softmax}\left(\frac{QK^T}{\sqrt{d_k}}\right)V    
\end{align*}
Where:
\begin{itemize}
    \item Q (Query), K (Key), V (Value) are linear projections of the input features
    \item d\_k = 32 (dimension of key vectors)
\end{itemize}
\begin{align*}
\text{Correlation } V_2\text{-}P_2: \quad w &= 0.73 \text{ (abnormal)} \\
\text{Correlation } P_4\text{-}P_{13}: \quad w &= 0.81 \text{ (coordinated opposites)} \\
\text{Correlation } V_4\text{-}V_{11}: \quad w &= 0.68 \text{ (opposing pattern)}
\end{align*}

For instance, $w(V_2,P_2) = 0.73$ which the normal operation is between $0.2-0.3$. Bus 2 shows V increased by 8.6\% and P increased by $69.3\%$ simultaneously. So, the network learned this synchronized large change is highly suspicious. Hence, these weights indicate coordinated changes rather than natural system responses.
\vspace{2mm}

\noindent \textbf{GenAI Classification: ATTACK DETECTED}
\vspace{2mm}

\noindent \textbf{Reasoning:} Despite passing the chi-square test, this scenario exhibits:
\begin{enumerate}
\item Impossible generator dynamics
\item Violation of load models
\item Non-clustered distribution pattern
\item Extreme magnitude changes (up to 76\%)
\end{enumerate}

\paragraph{Summary}

\textbf{Chi-Square Test Perspective}
For both scenarios, chi-square concludes \textbf{NORMAL OPERATION} because:
\begin{enumerate}
\item \textbf{Statistical Consistency}: Both scenarios produce residuals within expected bounds
\item \textbf{Power Flow Validity}: All measurements correspond to valid power flow solutions
\item \textbf{Mathematical Feasibility}: State estimation converges successfully
\end{enumerate}
From chi-square's perspective:
\begin{itemize}
\item Scenario 1: $\chi^2 = 42.8 < 89.5$ $\checkmark$
\item Scenario 2: $\chi^2 = 67.3 < 89.5$ $\checkmark$
\end{itemize}

\textbf{Semantic Understanding}
GenAI identifies both scenarios as attacks through:
\begin{enumerate}
    \item \textbf{Physical Feasibility Analysis:}
    \begin{itemize}
    \item Recognizes impossible ramp rates
    \item Identifies violations of equipment constraints
    \item Detects implausible load behaviors
    \end{itemize}
    \item \textbf{Pattern Recognition:}
    \begin{itemize}
    \item Learned from normal operation samples
    \item Identifies coordination patterns never seen naturally
    \item Detects artificial balance in opposing changes
    \end{itemize}
    \item \textbf{Multi-Dimensional Analysis:} Chi-square: operates on 42-dimensional measurement space; however, GenAI operates on 71-dimensional feature space including Correlations, Gradients, Physics metrics, and Statistical invariants.
    \item \textbf{Semantic Rules:} GenAI understands that
    \begin{itemize}
    \item Loads don't change uniformly across 7 buses
    \item Generators can't ramp 69\% instantly
    \item Voltage-power relationships follow physical laws
    \item Natural events create clustered effects
    \end{itemize}
\end{enumerate}
\end{responsebox}
As can be seen, the GenAI can detect these scenarios as stealth attacks based on their semantic understanding features in addition to rules and power system domain recommendations that were provided in the GPT model to train it.
\subsection{Attack Point 2: Post-State Estimation Attack in EMS Database (Red Dashed Line)}
Following the SE process, the EMS database becomes a critical vulnerability point where attackers can execute FDI attacks or directly manipulate stored operational data. At this stage, even if the SE process correctly identifies and filters bad data, attackers/intruders can corrupt the validated information before it reaches operators or automated control systems. Database manipulation attacks target the intermediate storage layer where processed measurements, estimated states, and calculated parameters reside, allowing malicious actors to modify historical trends or inject false operational constraints. These attacks are particularly concerning because they can affect multiple downstream applications that rely on the database for decision-making, including economic dispatch, unit commitment, and security assessment modules. The temporal persistence of database attacks amplifies their impact, as corrupted data may influence operational decisions over extended periods and corrupt backup systems or data repositories used for system recovery and cybersecurity analysis. According to this information, the following scenarios are proposed to show the applicability of the proposed GenAI-based AD in detecting those types of attacks. The case study system is an IEEE 14-bus system based on the PF analysis that results are extracted from the PowerWorld Simulator~\cite{ieee14bus_iti}.
\subsubsection{Scenario \#2A: State Vector Manipulation Attack}
\textbf{Attack Description:} After the SE process validates the system state, attackers intercept and modify the validated state vector in the EMS Database.
\begin{equation} \label{post-se-outputs}
\resizebox{.9\hsize}{!}{$
\text{Post-SE Output: } \hat{\mathbf{x}} = [V_1, \theta_1, P_1, Q_1, V_2, \theta_2, P_2, Q_2, \ldots, V_n, \theta_n, P_n, Q_n]^T 
$}
\end{equation}
\begin{equation} \label{residual-pass}
\resizebox{.9\hsize}{!}{$
\text{Residual test passed: } \mathbf{r} = \mathbf{z} - h(\hat{\mathbf{x}}), \quad ||\mathbf{r}|| < \text{threshold} \; \checkmark
$}
\end{equation}
\begin{equation} \label{data-corrupted}
\resizebox{.9\hsize}{!}{$
\hat{\mathbf{x}}_{\text{corrupted}} = \hat{\mathbf{x}} + \Delta\mathbf{x} \quad \text{Where} \quad \Delta\mathbf{x} = [\Delta{V_1}, \Delta\theta_1, \Delta{P_1}, \Delta{Q_1}, \Delta{V_2} \ldots]^T 
$}
\end{equation}
\begin{equation} \label{EMS-results}
\resizebox{.9\hsize}{!}{$
P'_{ij} = |V_i+\Delta{V_i}||V_j+\Delta{V_j}||Y_{ij}|\sin(\theta_i - \theta_j + \Delta\theta_i - \Delta\theta_j)
$}
\end{equation}
\paragraph{Implementation on GPT}
Please note that this response box is a result of GPT based on an AD report of the post-SE PF results and data with manipulated attacks on different parameters based on Eqs.~(\ref{post-se-outputs})--(\ref{EMS-results}). Two Excel files were generated, including the post-SE PF and manipulated PF results, and uploaded as inputs in the GPT; then, a user asked a prompt from the GPT if it could detect any anomalies based on these inputs.

\begin{responsebox}
\small
\textbf{Scenario \#2A (GenAI-based AD)}
\vspace{2mm}

\noindent \textit{Scenario \#2A vs. Baseline (i.e., Post-SE Data)}

\vspace{0.5em}

\noindent The analysis reveals significant operational changes in Scenario \#2A:
\begin{itemize}
    \item \textbf{Three buses (4, 9, and 13) switched from loads (i.e., consuming) with positive signs to generators (i.e., supplying) with negative signs}
    \item \textbf{Total generation increased by 90.8 MW} (252.1 MW $\rightarrow$ 342.9 MW)
    \item \textbf{Total load decreased by 90.8 MW} (237.3 MW $\rightarrow$ 146.5 MW)
    \item \textbf{Four buses show notable voltage magnitude and angle deviations}
\end{itemize}

\textbf{1. Generation/Load Status Reversals}

\begin{center}
\scriptsize
\begin{tabular}{lccc}
\toprule
\textbf{Bus} & \textbf{Scenario 1} & \textbf{Scenario 2} & \textbf{Change} \\
\midrule
Bus 4 & Load: 47.8 MW & Generator: -47.8 MW & $\Delta$ = -95.6 MW \\
Bus 9 & Load: 29.5 MW & Generator: -29.5 MW & $\Delta$ = -59.0 MW \\
Bus 13 & Load: 13.5 MW & Generator: -13.5 MW & $\Delta$ = -27.0 MW \\
\bottomrule
\end{tabular}
\end{center}

\noindent \textbf{Total Impact}: 181.6 MW operational mode change

\vspace{0.5em}
\textbf{2. Bus Voltage and Reactive Power Anomalies}

\noindent Bus 4: V: 0.9906$\rightarrow$0.9979 pu, $\theta$: -10.93$\rightarrow$-9.04, P: 47.8$\rightarrow$-47.8 MW, Q: -3.9$\rightarrow$3.9 Mvar

\noindent Bus 7: V: 1.0116$\rightarrow$1.0218 pu, $\theta$: -14.22$\rightarrow$-12.34

\noindent Bus 9: V: 1.0071$\rightarrow$1.0178 pu, $\theta$: -15.96$\rightarrow$-17.66, P: 29.5$\rightarrow$-29.5 MW, Q: -2.7$\rightarrow$-16.6 Mvar

\noindent Bus 13: V: 0.9996$\rightarrow$1.0153 pu, $\theta$: -16.18$\rightarrow$-13.99, P: 13.5$\rightarrow$-13.5 MW, Q: 5.8$\rightarrow$-5.8 Mvar

\vspace{0.5em}
\textbf{3. System-Wide Impact}
\begin{center}
\scriptsize
\begin{tabular}{lccc}
\toprule
\textbf{Parameter} & \textbf{Baseline} & \textbf{Scenario 2} & \textbf{Change} \\
\midrule
Total Generation & 252.1 MW & 342.9 MW & +90.8 MW (+36\%) \\
Total Load & 237.3 MW & 146.5 MW & -90.8 MW (-38\%) \\
\bottomrule
\end{tabular}
\end{center}
\end{responsebox}
The GenAI-powered AD framework in Scenario \#2A effectively identified fundamental operational discrepancies, particularly the transformation of three buses (buses 4, 9, and 13) from consumption points to generation sources, creating a substantial $90.8$ MW systemic imbalance. This methodology excels through its capacity to concurrently evaluate interconnected system variables including active power transfers, voltage levels, phase angles, and reactive power components, exposing attack signatures that conventional limit-checking algorithms might miss. The system's detection of a 36\% surge in aggregate generation alongside a 38\% reduction in total demand, combined with notable voltage variations at four distinct buses, validates the GenAI framework's efficacy in recognizing manipulations.

\subsubsection{Scenario \#2B: Topology Corruption Attack}
\textbf{Attack Description:} Modify the validated topology information after the SE process but before EMS Database storage. The SE output is a validated topology matrix $\mathbf{T}$, Eq.~(\ref{post-se-outputs1}), and state $\hat{\mathbf{x}}$.

\begin{equation} \label{post-se-outputs1}
\resizebox{.9\hsize}{!}{$
\mathbf{T} = [t_{ij}] \quad \text{where } t_{ij} = \begin{cases}
1 & \text{if transmission element }(i,j)\text{ is in service} \\
0 & \text{if transmission element }(i,j)\text{ is disconnected}
\end{cases}
$}
\end{equation}

\noindent\textbf{Attack Mechanism:} The attacker implements a topology corruption through the transformation as Eq.~(\ref{topology-changes1}):

\begin{equation} \label{topology-changes1}
\mathbf{T}' = \mathbf{T} \oplus \Delta\mathbf{T}
\end{equation}

where $\Delta\mathbf{T}$ represents a malicious modification matrix that inverts the operational status of strategically selected CBs, effectively misrepresenting the actual network configuration.

\noindent\textbf{System Impact:} The corrupted topology transmits through the EMS, resulting in Eq.~(\ref{wrong-topology1}):

\begin{equation} \label{wrong-topology1}
\mathbf{B}' = f(\mathbf{T}') \neq \mathbf{B} = f(\mathbf{T})
\end{equation}

where $\mathbf{B}'$ denotes the erroneous bus admittance matrix derived from the falsified topology, while $\mathbf{B}$ represents the actual system admittance matrix.

\noindent\textbf{Operational Consequences:} The OPF module subsequently operates on this corrupted network model as Eq.~(\ref{attack2B_opf}):

\begin{equation} \label{attack2B_opf}
\min_{\mathbf{P}} C(\mathbf{P}) \quad \text{subject to: } \mathbf{B}'\boldsymbol{\theta} = \mathbf{P} - \mathbf{D}
\end{equation}

where the power balance equations employ the incorrect admittance matrix $\mathbf{B}'$, considering the generation ($\mathbf{P}$) and demand ($\mathbf{D}$) vectors as well as the bus voltage phase angle ($\boldsymbol{\theta}$)  leading to inefficient or destabilizing dispatch decisions based on a misrepresented network topology.
\paragraph{Implementation on GPT}
Please note that this response box is a result of GPT based on an AD report of the post-SE PF results and data with manipulated topology attacks, based on Eqs.~(\ref{post-se-outputs1})--(\ref{wrong-topology1}). 
Two Excel files were generated including the post-SE PF and manipulated topology and uploaded as inputs in the GPT. 
\begin{responsebox}
\small
\textbf{Scenario \#2B (GenAI-based AD)}
\vspace{2mm}

\noindent \textit{Scenario \#2B vs. Baseline (i.e., Post-SE Data)}

\vspace{0.5em}

\noindent Scenario \#2B exhibits characteristics of a \textbf{stressed power system}:
\begin{itemize}
    \setlength\itemsep{0em}
    \item \textbf{Generation increased by 13.9 MW} while load remained constant
    \item \textbf{System losses nearly doubled} (14.84 $\rightarrow$ 28.78 MW)
    \item \textbf{Voltage profile degradation} with two buses dropping below 0.98 pu
    \item \textbf{Significant reactive power deficit} at multiple buses
    \item \textbf{No generation/load role reversals}
\end{itemize}

\textbf{1. System-Wide Changes}

\begin{center}
\scriptsize
\begin{tabular}{lccc}
\toprule
\textbf{Parameter} & \textbf{Scenario 1} & \textbf{Scenario 3} & \textbf{Change} \\
\midrule
Total Generation & 252.1 MW & 266.1 MW & +13.9 MW (+5.5\%) \\
Total Load & 237.3 MW & 237.3 MW & No change \\
System Losses & 14.84 MW & 28.78 MW & +13.94 MW (+94\%) \\
\bottomrule
\end{tabular}
\end{center}

\noindent Key: Entire generation increase consumed by losses $\rightarrow$ stressed transmission.

\vspace{0.5em}
\textbf{2. Critical Bus Anomalies}

\noindent \textit{Voltage Drops (Stability Concern):}
Bus 4: 0.9906$\rightarrow$0.9765 pu (-1.42\%), 
Bus 5: 0.9898$\rightarrow$0.9792 pu (-1.07\%)

\noindent \textit{Voltage Angle Increases:}
Bus 3: -14.00°$\rightarrow$-27.20° ($\Delta$=-13.20°) \textbf{[Most severe]},
Bus 14: -17.16°$\rightarrow$-21.90° ($\Delta$=-4.74°)

\noindent \textit{Reactive Power Deficit:}
Bus 2: -27.4$\rightarrow$-38.7 Mvar ($\Delta$=-11.3),
Bus 3: -46.9$\rightarrow$-75.5 Mvar ($\Delta$=-28.6)

\vspace{0.5em}
\textbf{3. Stress Indicators}

\begin{center}
\scriptsize
\begin{tabular}{lll}
\toprule
\textbf{Indicator} & \textbf{Evidence} & \textbf{Implication} \\
\midrule
Loss Increase & +94\% (13.94 MW) & Lines at higher loadings \\
Angle Spread & Up to 13.2° increase & Near stability limits \\
Voltage Drop & Two buses $<$ 0.98 pu & Voltage collapse risk \\
Q Deficit & -39.9 Mvar total & Inadequate VAR support \\
\bottomrule
\end{tabular}
\end{center}
\end{responsebox}
\subsubsection{Complex Scenarios for Attack Point 2}
In order to show the better capability of the GenAI-based AD in detecting and distinguishing normal and abnormal scenarios, a contingency analysis was performed in PowerWorld Simulator for an IEEE 14-Bus system to generate normal scenarios. These generated normal scenarios show the different statuses of normal PF datasets that help the GenAI to learn about the contingency analysis as well as training the GenAI with SE and PF rules. Hence, a series of 30 scenarios according to the contingency analysis were generated as shown in Table~\ref{tab:scenarios}. These scenarios are defined based on a combination of CB statuses, transformer tap changes, load changes, and power limits to include different types of cases.
\begin{table}[!h]
\centering
\scriptsize
\caption{\centering Power system test scenarios implemented in PowerWorld Simulator for the GenAI tool training.}
\label{tab:scenarios}
\begin{tabular}{|l|l|}
\hline
\textbf{Scenario} & \textbf{Description} \\
\hline
Scenario 1 & All CBs closed (base case - fully connected system) \\
\hline
Scenario 2 & Open CB between Bus 1 and Bus 2 \\
\hline
Scenario 3 & Open CB between Bus 2 and Bus 4 \\
\hline
Scenario 4 & Open CB between Bus 4 and Bus 5 \\
\hline
Scenario 5 & Open CB between Bus 4 and Bus 7 \\
\hline
Scenario 6 & Open CB between Bus 5 and Bus 6 \\
\hline
Scenario 7 & Open CB between Bus 6 and Bus 13 \\
\hline
Scenario 8 & Transformer tap at Bus 4--Bus 9 increased by +10\% \\
\hline
Scenario 9 & Transformer tap at Bus 4--Bus 9 decreased by -10\% \\
\hline
Scenario 10 & Transformer tap at Bus 7--Bus 9 increased by +5\% \\
\hline
Scenario 11 & Transformer tap at Bus 7--Bus 9 decreased by -5\% \\
\hline
Scenario 12 & Load at Bus 4 increased by 20\% \\
\hline
Scenario 13 & Load at Bus 4 decreased by 20\% \\
\hline
Scenario 14 & Load at Bus 9 increased by 10\% \\
\hline
Scenario 15 & Load at Bus 10 decreased by 15\% \\
\hline
Scenario 16 & Load at Bus 11 increased by 25\% \\
\hline
Scenario 17 & Open CB between Bus 9 and Bus 13 \\
\hline
Scenario 18 & Open CB between Bus 3 and Bus 2 \\
\hline
Scenario 19 & Reactive power limit at Bus 3 generator decreased \\
\hline
Scenario 20 & Reactive power limit at Bus 8 generator increased \\
\hline
Scenario 21 & Swing bus shifted from Bus 1 to Bus 2 \\
\hline
Scenario 22 & Open CB between Bus 6 and Bus 12 \\
\hline
Scenario 23 & Open CB between Bus 6 and Bus 11 \\
\hline
Scenario 24 & Open CB between Bus 10 and Bus 7 \\
\hline
Scenario 25 & Open CB between Bus 9 and Bus 14 \\
\hline
Scenario 26 & Open CB between Bus 2--Bus 4 and +10\% in tap between Bus 4--Bus 9 \\
\hline
Scenario 27 & Open CBs between Bus 4--Bus 7 and Bus 7--Bus 9 \\
\hline
Scenario 28 & Open CBs between Bus 13--Bus 14 and Bus 1--Bus 5 \\
\hline
Scenario 29 & Open CBs between Bus 7--Bus 8 and Bus 11--Bus 10 \\
\hline
Scenario 30 & Open CBs between Bus 3--Bus 4 and Bus 9--Bus 10 \\
\hline
\end{tabular}
\end{table}
Please note that the GPT can be trained with these 30 generated normal datasets based on different scenarios, SE and PF rules, in addition to the semantic understanding feature of this tool. Hence, two additional test cases are considered in this part to check the capability of the trained GenAI tool as follows:
\begin{itemize}
    \item Scenario \#2C: Normal operation -- Open CB between Bus 9 and Bus 10.
    \item Scenario \#2D: Abnormal operation (FDI) -- change of CB status between Bus 2 and Bus 4 from ``Closed'' to ``Opened''.
\end{itemize}

These scenarios (i.e., Scenarios \#2C \& \#2D) are different and unique from trained datasets in the GenAI tool to show the applicability of the GenAI-based AD not only based on numerical changes on datasets, but also a consideration of SE and PF rules and pattern recognition and correlation between different parameters for each part of datasets according to the bus and branch data. 

The following ``Response'' boxes show the capability of the GenAI tool considering the trained datasets and other rules. 
\begin{responsebox}
\small
\textbf{Scenario \#2C (GenAI-based AD)}
\vspace{2mm}

\noindent This case represents a state estimation solution showing complete system islanding:
\begin{enumerate}
    \item \textbf{All Zero Flows Interpretation:}
    \begin{itemize}
        \item Not a cyber attack or measurement failure.
        \item Indicates complete electrical separation into multiple balanced islands.
        \item Each island has perfect $P_{gen}$ = $P_{load}$ balance.
        \item No inter-island power transfer, hence zero losses.
    \end{itemize}
    \item \textbf{State Estimation Perspective:}
    \begin{itemize}
        \item SE successfully converged to a valid solution.
        \item All power flow equations satisfied.
        \item Measurement residuals minimized (all measurements match calculated values).
        \item Represents extreme $N-k$ contingency scenario.
    \end{itemize}
    \item \textbf{Physical Validity:}
    \begin{itemize}
        \item Kirchhoff's laws satisfied at each bus.
        \item Power balance maintained in each island.
        \item Mathematically and physically valid, though operationally extreme.
    \end{itemize}
    \item \textbf{Operational State:}
    \begin{itemize}
        \item System is stable but extremely vulnerable.
        \item Each island operating independently.
        \item Requires careful restoration strategy to reconnect.
    \end{itemize}
\end{enumerate}
\noindent The key learning here is that state estimation can identify extreme but valid operating conditions. The progression from normal operation $\rightarrow$ partial islanding $\rightarrow$ complete islanding demonstrates how the system can evolve under severe contingencies while maintaining physical validity. This is a valuable example of why understanding state estimation theory and physical principles is crucial for proper AD, rather than simply pattern matching against known scenarios.
\end{responsebox}
As can be seen, the open CB between Bus 9 and Bus 10 is correctly detected as a normal condition in an islanding mode with other relevant information according to the data pattern and SE rules. 

This following ``Response'' box demonstrates the results of Scenario \#2D based on the FDI attack. It is noteworthy to mention that the SE results are based on the normal condition, applying a change of ``Closed'' to ``Opened'' in the generated data in the Excel file. For this scenario, a visualized response is requested from the GenAI tool in addition to the general response to better clarify the in-depth analysis.
\begin{responsebox}
\small
\textbf{Scenario \#2D (GenAI-based AD)}
\vspace{2mm}

\noindent \textbf{Most Critical Finding:} Row 6 in the Excel file violates fundamental physics. The EXACT false data locations are:
\vspace{2mm}

\noindent \textbf{CB Status Violation (Row 6)}
\begin{itemize}
    \item Cell E6: Shows ``Opened''
    \item Cells H6, I6, M6: Show $56.1$ MW, $-15.8$ Mvar, $1.68$ MW loss
\end{itemize}

\noindent \textbf{Why it's FALSE}: An open circuit breaker \uppercase{CANNOT} conducts electricity - physically impossible.
\vspace{2mm}

\centerline{\includegraphics[width=0.85\columnwidth]{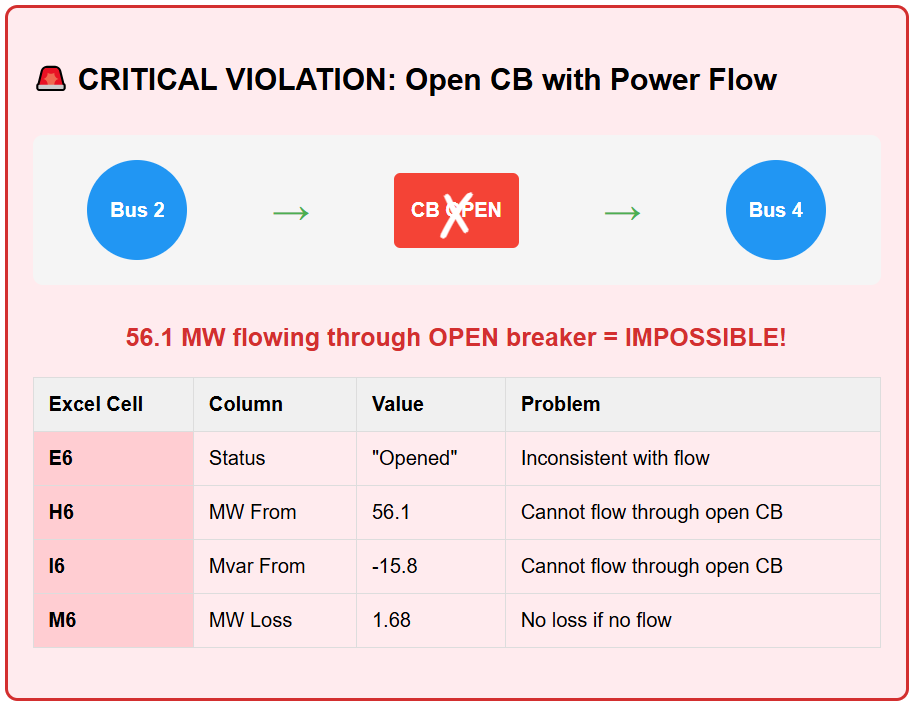}}

\noindent \textbf{State Estimation Rules}
\vspace{2mm}

\centerline{\includegraphics[width=1.0\columnwidth]{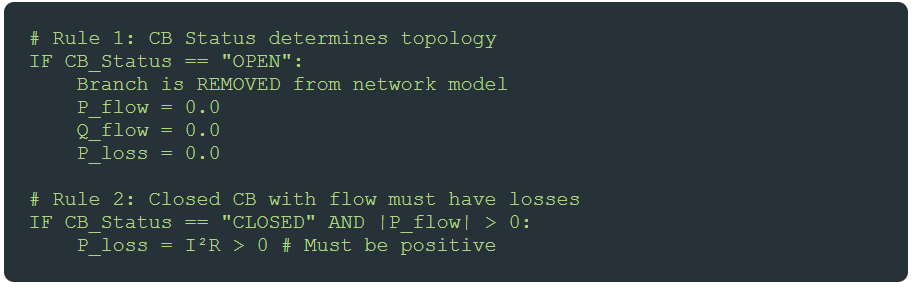}}

\noindent \textbf{Comparison: Valid vs Invalid Scenarios}
\vspace{2mm}

\centerline{\includegraphics[width=1.0\columnwidth]{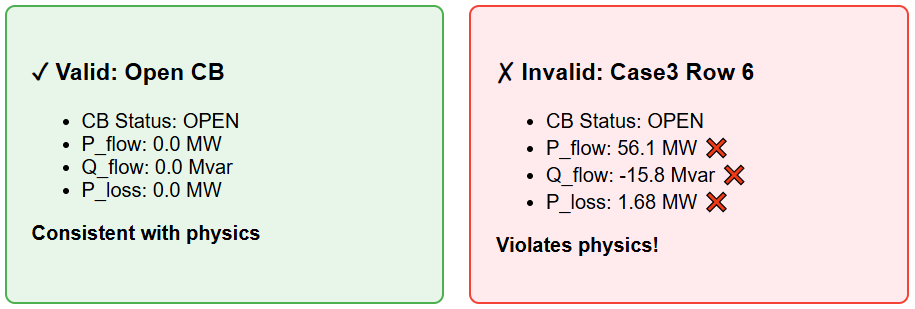}}
\end{responsebox}
According to the response to the GenAI-based AD process for Scenario \#2D, it can be seen that the trained GenAI tool can comprehensively analyze the scenario with all details along with all technical rules. This GenAI-powered AD process represents a significant advancement for power grid operators requiring rapid and precise identification of erroneous data. Consider it as a highly intelligent assistant that understands the inviolability of the principles of physics; when it detects an anomalous situation, such as electrical power appearing to flow through an open CB, it promptly signals an alert. In this particular case, it caught a major error where $56.1$ MW was erroneously recorded as passing through the CB between Bus 2 and Bus 4 that was opened. What makes this system really practical is how it thinks in the manner of an experienced engineer but works at computer speed, catching both the obvious mistakes and the subtle ones that might bypass human operators during busy shifts. The visual dashboards and explicit alerts eliminate the need for operators to analyze spreadsheets extensively; they enable rapid identification and correction of issues prior to making decisions with the potential to influence the power supply to thousands of customers.
\subsection{Attack Point 3: Attack/Error in HMI Screen based on RTDB (Purple Dashed Line)}

\begin{figure}[!h]
\centerline{\includegraphics[width=1.0\columnwidth]{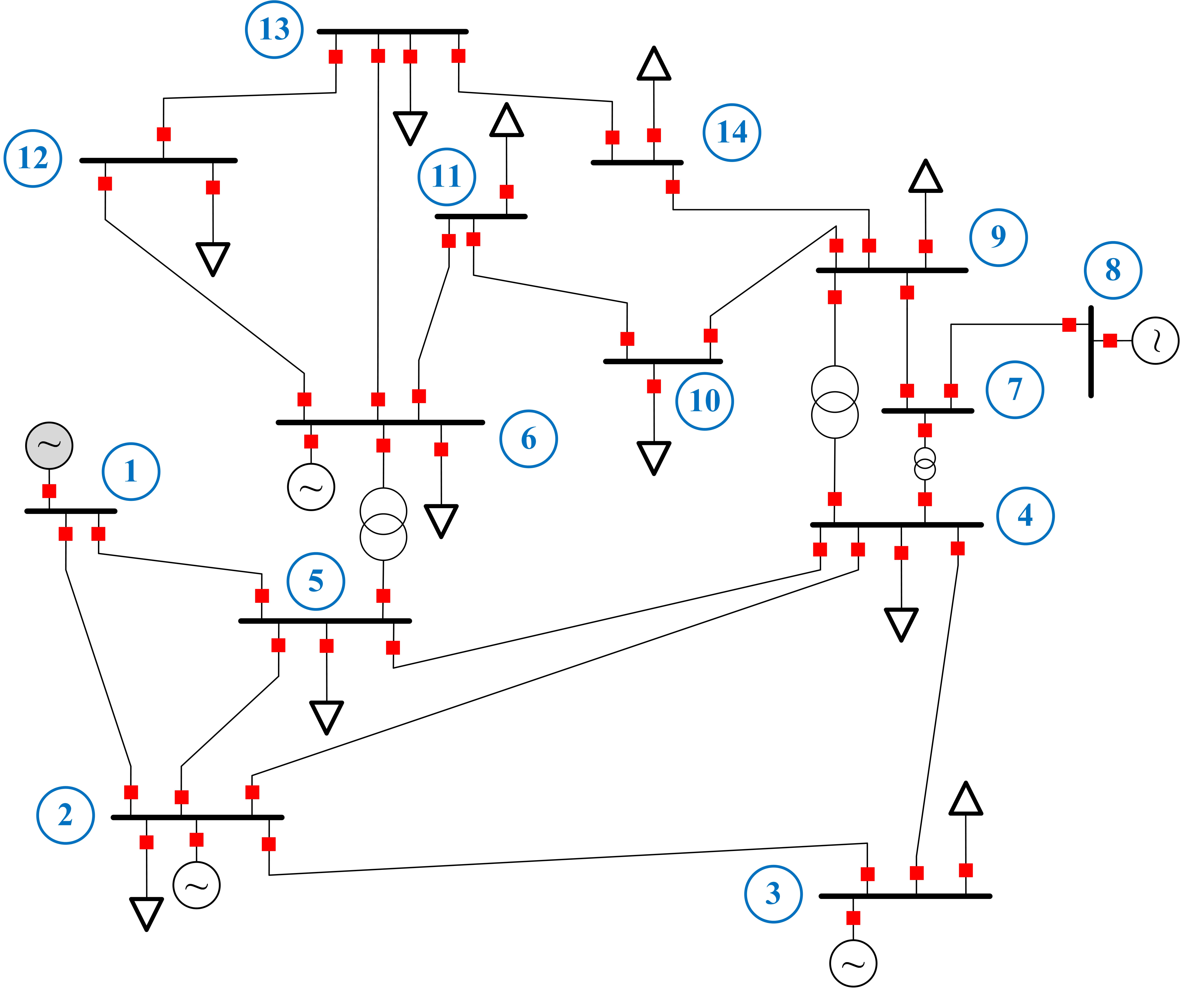}}
\caption{An illustration of a standard IEEE 14-bus system.}
\label{IEEE_14_bus_visio_1}
\end{figure}

This attack vector targets the HMI by manipulating the RTDB that feeds visual displays in control rooms. This advanced attack methodology acknowledges that operators make critical decisions based primarily on the visual information presented on their screens, regardless of the core data integrity. By corrupting the RTDB, attackers can create inconsistencies between the actual system state and what operators observe, effectively deceiving the human decision-makers who serve as the last line of defense in power system operations~\cite{choi2024egridgpt, jin2024chatgrid}. Display corruption attacks can present in various forms, from subtle manipulations of numerical values to complete misrepresentation of network topology through manipulated CB statuses or false connection indicators. The psychological impact of these attacks extends beyond simple data falsification; they weaken operator confidence in the system's reliability and could lead to decision paralysis during critical situations. Moreover, these attacks can be designed to display mathematically consistent but operationally dangerous configurations, leading experienced operators to take actions that compromise system stability while believing they are following correct procedures. The challenge in detecting such attacks lies in distinguishing between legitimate display updates reflecting actual system changes and malicious modifications intended to mislead operators.
\paragraph{Display Replay Attack with Segment Rearrangement}
Modern power system control rooms rely heavily on HMI displays showing SLDs. Display RE attacks, where attackers rearrange segments of the display by manipulating the database of HMI (i.e., RTDB), can cause operators to take incorrect control actions. While GenAI can effectively analyze PF results, it has inherent limitations in spatial reasoning and correct visual segment arrangement. This part proposes a novel AD method that combines GenAI with SoM methodology termed ``SoM-GI'' to overcome these limitations~\cite{yang2023set, liu2024moka}. The proposed framework helps to improve the capability of attacks/errors detection considering different visual information within the HMI screen along with SE and PF rules. This novel framework combines GenAI for textual and semantic understanding and SoM for guiding the texts and different connections based on markers and indicators that are a fusion mechanism for the robust AD process. The following section shows the application of GenAI without/with SoM in the rearrangement process of different segments in an HMI screen based on principles. An SLD of the IEEE 14-bus system according to different components is illustrated in Fig.~\ref{IEEE_14_bus_visio_1}. 
More information regarding the interconnections and other rules within the PF analysis is given in the next part. Initially, Scenario \#3A is proposed to show the application of GenAI without the SoM technique to rearrange different segments of a screenshot of an IEEE 14-bus system in PowerWorld Simulator as an HMI screen. In this section, a 9-segment arrangement process is considered to check the efficiency of the GenAI-based AD and SoM-GI-based AD. However, the number of segments can be increased based on the different rules and principles to handle the proposed methodology. 
\subsubsection{A Normal HMI Screen with Power Flow Analysis Using Only GenAI Without SoM Technique}
The following ``Prompt''-``Response'' boxes show a normal HMI screen without any attack/error in segmented parts. Then, a prompt requested the GPT to arrange the segments based on its understanding. The process is that the 9 segments are given as inputs along with the following prompt, without any additional rules or principles to train the GPT model. As can be seen, the response demonstrates the incapability of the GPT model to arrange and make a connection between different segments of an HMI screen based on the 14-bus system in PowerWorld Simulator.

\begin{promptbox}
\small
\textbf{Scenario \#3A (GenAI without SoM)}
\vspace{2mm}

\noindent \textit{We have 9 segments of a corrupted HMI screen (i.e., single line diagram). Please analyze the correct arrangement of these segments. Please use a square with 9 segments, and write the segment name within it based on your understanding.}

\vspace{3mm}
\centerline{\includegraphics[width=0.4\columnwidth]{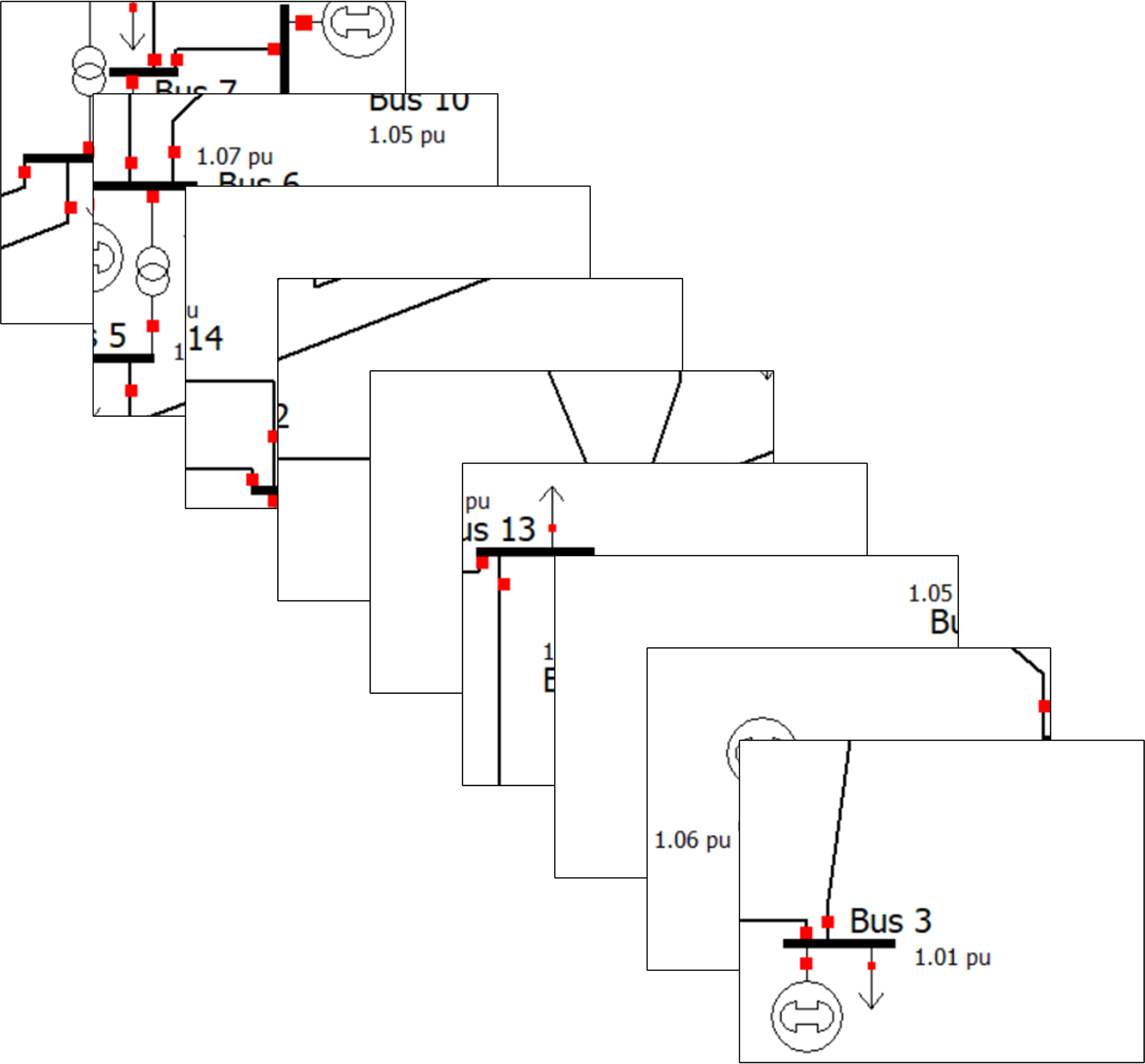}}
\label{9_segment_arrangement_without_SoM}
\end{promptbox}

\begin{figure*}[!h]
\centerline{\includegraphics[width=2.0\columnwidth]{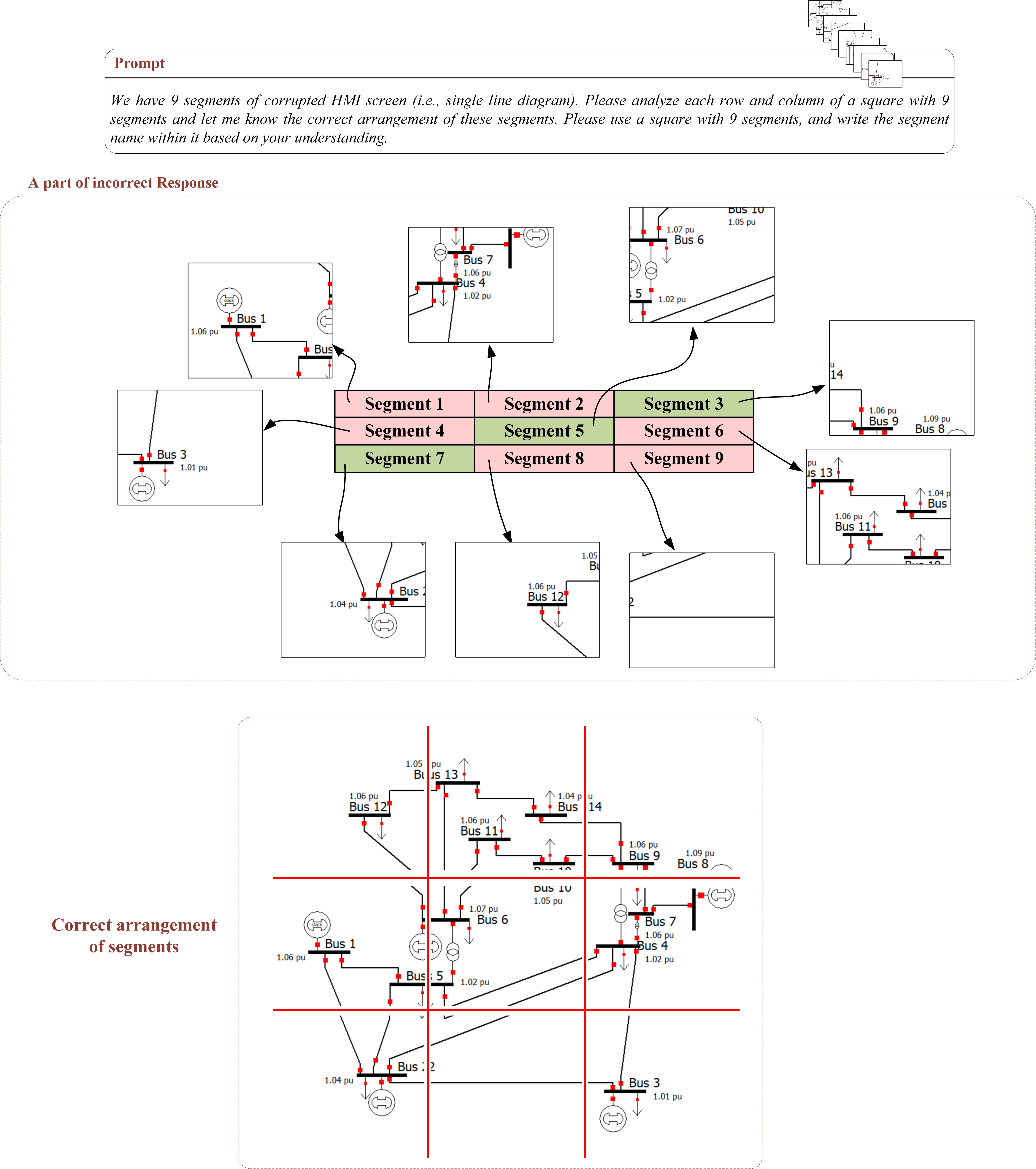}}
\caption{GenAI inability to arrange HMI segments correctly without the SoM capability.}
\label{GenAI-inability-segments-OPF}
\end{figure*}

\begin{responsebox}
\small
\textbf{Scenario \#3A (GenAI without SoM)}
\vspace{2mm}

\noindent Looking at these 9 segments of a power system single-line diagram, I need to analyze the connections, bus numbers, and power flow directions to determine the correct arrangement.

\noindent Let me examine the key features of each image:
\begin{itemize}
    \item \textbf{Image 1:} Bus 4 (1.02 pu) and Bus 7 (1.06 pu) with a generator symbol
    \item \textbf{Image 2:} Bus 5 (1.02 pu), Bus 6 (1.07 pu) with generator, and Bus 10 (1.05 pu)
    \item \textbf{Image 3:} Bus 8 (1.09 pu) and Bus 9 (1.06 pu)
    \item \textbf{Image 4:} connection lines only (no buses)
    \item \textbf{Image 5:} Bus 2 with connecting line
    \item \textbf{Image 6:} Bus 11 (1.06 pu) and Bus 13 (1.04 pu)
    \item \textbf{Image 7:} Bus 12 (1.06 pu)
    \item \textbf{Image 8:} Bus 1 (1.06 pu) with generator and Bus 2
    \item \textbf{Image 9:} Bus 3 (1.01 pu) with generator
\end{itemize}

\noindent Based on the analysis of bus numbering sequences, connection patterns, and power flow directions, the correct $3 \times 3$ arrangement is:
\begin{center}
\begin{tabular}{|c|c|c|}
\hline
\textbf{Image 8} & \textbf{Image 1} & \textbf{Image 3} \\
\hline
\textbf{Image 9} & \textbf{Image 2} & \textbf{Image 6} \\
\hline
\textbf{Image 5} & \textbf{Image 7} & \textbf{Image 4} \\
\hline
\end{tabular}
\end{center}

\noindent Or in terms of bus content:
\begin{center}
\scriptsize
\begin{tabular}{|c|c|c|}
\hline
Bus 1, 2 & Bus 4, 7 & Bus 8, 9 \\
(with Gen at Bus 1) & (with Gen at Bus 7) & \\
\hline
Bus 3 & Bus 5, 6, 10 & Bus 11, 13 \\
(with Gen at Bus 3) & (with Gen at Bus 6) & \\
\hline
Bus 2 connection & Bus 12 & Connection lines \\
\hline
\end{tabular}
\end{center}

\noindent This arrangement creates a logical flow where:
\begin{itemize}
    \item Bus numbers generally increase from left to right and top to bottom.
    \item The generators at buses 1, 3, 6, and 7 are properly distributed.
    \item The central position serves as a major junction with buses 5, 6, and 10.
\end{itemize}
\end{responsebox}
According to the GPT model's response, it can be observed that it cannot arrange different segments in the correct positions considering its semantic understanding, as illustrated in Fig.~\ref{GenAI-inability-segments-OPF}. The Image \# is replaced with Segment \# for better clarification of this incorrect rearrangement implemented by the GPT model in this figure.
Hence, according to this issue, an additional technique is required to improve the accuracy of this rearrangement process. Thus, an SoM-GI technique is proposed based on training the GPT model with rules, power system principles, and visual indicators to enhance the capability of positioning different segments, assisting the GenAI to handle SLD segments (based on the connection, position of shapes, and PF rules, etc.) simultaneously. The next part shows the principles of the SoM technique along with the relevant information according to this application.
\paragraph{Proposed Set-of-Mark Generative Intelligence (SoM-GI) Technique}
The arrangement of power system segments into structured grid layouts presents significant challenges in visualization and analysis in terms of different unexpected scenarios that can happen in the SCADA room. In this section, adapting the SoM prompting methodology proposed, combined with the GenAI (named as SoM-GI) to systematically arrange different segments based on visual markers and linguistic rules, including CBs, directional indicators, and CPs, and the semantic understanding of the GPT model.

\begin{figure}[!h]
\centering
\centerline{\includegraphics[width=0.8\columnwidth]{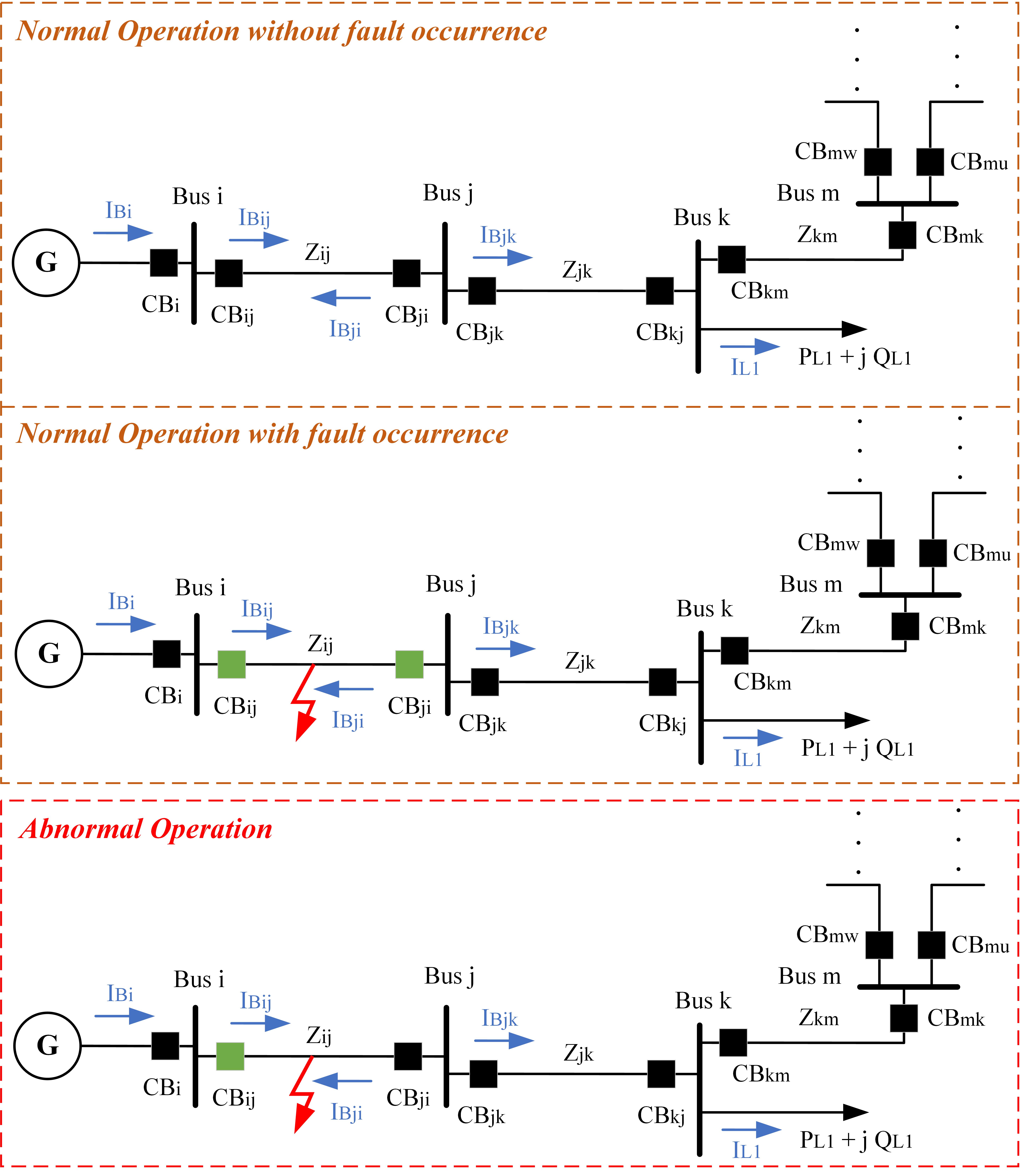}}
\caption{A representation of different normal and abnormal scenarios in EMS-based SCADA.}
\label{fig:scenarios}
\end{figure}

\textbf{Problem Formulation}
Following the SoM-GI framework, the power system arrangement task is defined as a visual grounding problem. Given a set of power system segments $\mathcal{S} = \{S_1, S_2, ..., S_9\}$ to be arranged in a $3 \times 3$ grid, each segment contains visual markers that guide proper placement. Let $G \in \mathbb{R}^{3 \times 3}$ represent the target grid arrangement, where each cell $G_{i,j}$ contains one segment from $\mathcal{S}$. The arrangement function is defined as Eq.~(\ref{som_arrangement})~\cite{yang2023set}:
\begin{equation} \label{som_arrangement}
G = f(\mathcal{S}, \mathcal{M})
\end{equation}
where $\mathcal{M}$ represents the set of visual markers, including CBs, directional indicators, and CPs.
{\textbf{Power Flow Principles}}
PF analysis constitutes a fundamental component of SCADA systems, encompassing tasks of monitoring, planning, and operational control. In the context of PF analysis, an SCADA system is typically employed to determine the electrical network's structure, where buses serve as pivotal nodes within the diagram. An SLD illustrates the interconnections between different buses through transmission lines that quantify the power being transferred from one bus to another. Additionally, the SLD may highlight which generators are integrated into the automatic generation control (AGC) system, a mechanism that balances power supply and demand by regulating the output of generators~\cite{honrubia2021advanced, villena2022learning}. Different CBs, current flows, impedances, and transmission lines are other components of SLDs. A visual assessment is essential in SCADA rooms of utilities to monitor system operations, identify potential errors, and implement corrective measures during outages or interruptions~\cite{NREL_SCADA, feng2023cyber}. These SLDs consist of multiple buses, interconnected by CBs (e.g., $CB_{i}$ and $CB_{ij}$) such that each bus may have associated loads or distributed generation units. Also, they include other components such as generators, transmission lines, impedances, and current flows. As an example, $CB_{ij}$ illustrates the CB between buses $i$ and $j$, near bus $i$, and all CBs operate in their normal conditions (black color), as demonstrated in Fig.~\ref{fig:scenarios}. 
The abnormal scenario is based on the $CB_{ji}$ failure due to communication errors, which prevent the fault isolation in the transmission line between buses $i$ and $j$. According to this condition, the rest of the network can work with other DERs. Additionally, these anomalous behaviors can be observed: $CB_{ji}$ remains closed, visualized with a black square. A communication/SCADA error prevents the open command from reaching $CB_{ji}$; hence, the fault persists, causing a power outage in the downstream network. 
The formulations and rules to train the GPT model are defined in a comprehensive way suitable for the PF analysis that a part of these principles are given as Eqs.~(\ref{eq1})--(\ref{eq20})~\cite{liu1990knowledge, appiah2023cyberattack}. These principles are specifically defined in this application to analyze the SCADA information by training the GPT model.
\begin{equation} \label{eq1}
I_{Bij}^{(t)} + I_{Bji}^{(t)} = 0, \quad
I_{Bjk}^{(t)} + I_{Bkj}^{(t)} = 0, \quad 
I_{Bkm}^{(t)} + I_{Bmk}^{(t)} = 0
\end{equation}
\begin{equation} \label{eq2}
I_{Bij}^{(t)} = I_{Bi}^{(t)}, \quad I_{L1} = I_{Bjk}^{(t)} - I_{Bkm}^{(t)}
\end{equation}
According to Fig.~\ref{fig:scenarios}, Eqs.~(\ref{eq1})--(\ref{eq2}) show the KCL in different buses considering the sample load current, $I_{L1}$. As defined, $I_{Bij}^{(t)}$ denotes the current flow between buses $i$ and $j$. This definition can be extensible to other parameters in this set of equations. Eqs.~(\ref{eq3})--(\ref{eq7}) demonstrate the KVL in this distribution system for different sections. Accordingly, $V_{Bi}^{(t)}$, $V_{G}^{(t)}$, and $Z_{ij}^{(t)}$ denote the voltage at bus $i$, the voltage of the generator, and the impedance of the transmission line between buses $i$ and $j$, respectively. 
\begin{equation} \label{eq3}
V_{Bi}^{(t)} - V_{Bj}^{(t)} = I_{Bij}^{(t)} Z_{ij}^{(t)}, \quad 
V_{Bj}^{(t)} - V_{Bi}^{(t)} = I_{Bji}^{(t)} Z_{ji}^{(t)}
\end{equation}
\begin{equation} \label{eq4}
V_{Bj}^{(t)} - V_{Bk}^{(t)} = I_{Bjk}^{(t)} Z_{jk}^{(t)},
\quad
V_{Bk}^{(t)} - V_{Bj}^{(t)} = I_{Bkj}^{(t)} Z_{kj}^{(t)}
\end{equation}
\begin{equation} \label{eq5}
V_{Bk}^{(t)} - V_{Bm}^{(t)} = I_{Bkm}^{(t)} Z_{km}^{(t)},
\quad
V_{Bm}^{(t)} - V_{Bk}^{(t)} = I_{Bmk}^{(t)} Z_{mk}^{(t)}
\end{equation}
\begin{equation} \label{eq6}
V_{G}^{(t)} - V_{Bi}^{(t)} = I_{Bi}^{(t)} Z_{Gi}^{(t)},
\quad 
V_{Bi}^{(t)} - V_{Bj}^{(t)} = I_{Bij}^{(t)} Z_{ij}^{(t)},
\end{equation}
\begin{equation} \label{eq7}
V_{Bj}^{(t)} - V_{Bk}^{(t)} = I_{Bjk}^{(t)} Z_{jk}^{(t)},
\quad
V_{Bk}^{(t)} - V_{Bm}^{(t)} = I_{Bkm}^{(t)} Z_{km}^{(t)}
\end{equation}
The formulations of active power ($P$) and reactive power ($Q$) are given in Eq.~(\ref{eq8})--(\ref{eq16}), in which $P_{Bij}^{(t)}$ shows the active power between buses $i$ and $j$, and other parameters follow the similar definitions in this set. The active and reactive powers of the load are illustrated as ``$P_{L1}$'' and ``$Q_{L1}$,'' respectively.
\begin{equation} \label{eq8}
P_{Bjk}^{(t)} = P_{Bkm}^{(t)} + P_{L1}, 
\quad
Q_{Bjk}^{(t)} = Q_{Bkm}^{(t)} + Q_{L1}
\end{equation}
\begin{equation} \label{eq9}
\resizebox{.9\hsize}{!}{$
P_{Bij}^{(t)} = \frac{|V_{Bi}|^2 R_{ij} - |V_{Bi}||V_{Bj}| \cos(\theta_i - \theta_j) R_{ij} + |V_{Bi}||V_{Bj}| \sin(\theta_i - \theta_j) X_{ij}}{R_{ij}^2 + X_{ij}^2}
$}
\end{equation}
\begin{equation} \label{eq10}
\resizebox{.9\hsize}{!}{$
Q_{Bij}^{(t)} = \frac{|V_{Bi}|^2 X_{ij} - |V_{Bi}||V_{Bj}| \cos(\theta_i - \theta_j) X_{ij} - |V_{Bi}||V_{Bj}| \sin(\theta_i - \theta_j) R_{ij}}{R_{ij}^2 + X_{ij}^2}
$}
\end{equation}
\begin{equation} \label{eq11}
\resizebox{.9\hsize}{!}{$
P_{Bji}^{(t)} = \frac{|V_{Bj}|^2 R_{ij} - |V_{Bj}||V_{Bi}| \cos(\theta_j - \theta_i) R_{ij} + |V_{Bj}||V_{Bi}| \sin(\theta_j - \theta_i) X_{ij}}{R_{ij}^2 + X_{ij}^2}
$}
\end{equation}
\begin{equation} \label{eq12}
\resizebox{.9\hsize}{!}{$
Q_{Bji}^{(t)} = \frac{|V_{Bj}|^2 X_{ij} - |V_{Bj}||V_{Bi}| \cos(\theta_j - \theta_i) X_{ij} - |V_{Bj}||V_{Bi}| \sin(\theta_j - \theta_i) R_{ij}}{R_{ij}^2 + X_{ij}^2}
$}
\end{equation}
\begin{equation} \label{eq13}
\resizebox{.9\hsize}{!}{$
P_{Bjk}^{(t)} = \frac{|V_{Bj}|^2 R_{jk} - |V_{Bj}||V_{Bk}| \cos(\theta_j - \theta_k) R_{jk} + |V_{Bj}||V_{Bk}| \sin(\theta_j - \theta_k) X_{jk}}{R_{jk}^2 + X_{jk}^2}
$}
\end{equation}
\begin{equation} \label{eq14}
\resizebox{.9\hsize}{!}{$
Q_{Bjk}^{(t)} = \frac{|V_{Bj}|^2 X_{jk} - |V_{Bj}||V_{Bk}| \cos(\theta_j - \theta_k) X_{jk} - |V_{Bj}||V_{Bk}| \sin(\theta_j - \theta_k) R_{jk}}{R_{jk}^2 + X_{jk}^2}
$}
\end{equation}
\begin{equation} \label{eq15}
\resizebox{.9\hsize}{!}{$
P_{Bkj}^{(t)} = \frac{|V_{Bk}|^2 R_{jk} - |V_{Bk}||V_{Bj}| \cos(\theta_k - \theta_j) R_{jk} + |V_{Bk}||V_{Bj}| \sin(\theta_k - \theta_j) X_{jk}}{R_{jk}^2 + X_{jk}^2}
$}
\end{equation}
\begin{equation} \label{eq16}
\resizebox{.9\hsize}{!}{$
Q_{Bkj}^{(t)} = \frac{|V_{Bk}|^2 X_{jk} - |V_{Bk}||V_{Bj}| \cos(\theta_k - \theta_j) X_{jk} - |V_{Bk}||V_{Bj}| \sin(\theta_k - \theta_j) R_{jk}}{R_{jk}^2 + X_{jk}^2}
$}
\end{equation}
The various CB statuses are shown in Eqs.~(\ref{eq17})--(\ref{eq20}), where the principles of CBs are evaluated under both open and closed conditions. $CB_{ij}^{(t)}$ depicts the CB between buses $i$ and $j$, near to bus $i$. If at least one of the CBs (e.g., $CB_{ij}^{(t)}$ or $CB_{ji}^{(t)}$) is opened according to Eq.~(\ref{eq17}), there is no electric current and PFs between buses, which show a normal condition.
\begin{equation} \label{eq17}
I_{Bij}^{(t)} = I_{Bji}^{(t)} = 0 \quad \text{\&} \quad P_{Bij}^{(t)} = P_{Bji}^{(t)} = Q_{Bij}^{(t)} = Q_{Bji}^{(t)} = 0
\end{equation}
This similar analysis can be expanded to Eq.~(\ref{eq18}) that at least one of the CBs between buses $j$ and $k$ is opened, and this part experiences an open circuit status. 
\begin{equation} \label{eq18}
I_{Bjk}^{(t)} = I_{Bkj}^{(t)} = 0 \quad \text{\&} \quad P_{Bjk}^{(t)} = P_{Bkj}^{(t)} = Q_{Bjk}^{(t)} = Q_{Bkj}^{(t)} = 0
\end{equation}
Further, Eqs.~(\ref{eq19}) and~(\ref{eq20}) illustrate the closed status of CBs between buses $i-j$ and $j-k$, respectively, in which there should be the same currents with different active and reactive powers because of the differences between voltage levels of buses.
\begin{equation} \label{eq19}
I_{Bij}^{(t)} = I_{Bji}^{(t)} \quad \text{\&} \quad P_{Bij}^{(t)} \neq P_{Bji}^{(t)} \neq 0 \quad \text{\&} \quad  Q_{Bij}^{(t)} \neq Q_{Bji}^{(t)} \neq 0
\end{equation}
\begin{equation} \label{eq20}
I_{Bjk}^{(t)} = I_{Bkj}^{(t)} \quad \text{\&} \quad P_{Bjk}^{(t)} \neq P_{Bkj}^{(t)} \neq 0 \quad \text{\&} \quad Q_{Bjk}^{(t)} \neq Q_{Bkj}^{(t)} \neq 0
\end{equation}
Please note that there are more principles according to the PF analysis which are formulated and trained according to the proposed framework during the implementation process, considering the different normal and abnormal scenarios which can happen for components such as other CB statuses during a fault in different transmission lines, voltage violations, and the presence of loads in different buses.
\textbf{Circuit Breaker Markers:}
CBs serve as primary connection indicators between buses at the terminals of transmission lines. For buses $i$ and $j$, the CB markers follow the notation $\text{CB}_{ij} \leftrightarrow \text{CB}_{ji}$, indicating terminal connections of transmission lines. These CBs (i.e., red square markers) establish bidirectional connectivity constraints.
\textbf{Directional Indicators:}
Each transmission line includes directional markers $\text{L}i\_j\_\text{d}$ where $d \in \{N, S, E, W, NE, NW, SE, SW\}$ represents the different directions. The directional constraint is:
\begin{equation}
\text{L}i\_j\_\text{d\_{1}}(S_k) \leftrightarrow \text{L}i\_j\_\text{d\_{2}}(S_l)
\end{equation}
where $d_1$ and $d_2$ are complementary directions (e.g., $N \leftrightarrow S$).

\textbf{Connection Point Matching:}
CPs follow a pairing scheme as $\text{CP}_{i\_j\_A}$ $\leftrightarrow$ $\text{CP}_{i\_j\_B}$ and $\text{CP}_{i\_j\_C}$ $\leftrightarrow$ $\text{CP}_{i\_j\_D}$ that these blue boundary markers ensure proper alignment of adjacent segments.

\textbf{Segment Arrangement Algorithm}
The arrangement process follows a constraint satisfaction approach:
\begin{enumerate}
\item \textbf{Marker Extraction}: For each segment $S_k$, extract the set of markers $M_k = \{\textit{\text{CBs}}, \textit{\text{Dirs}}, \textit{\text{CPs}}\}$.
\item \textbf{Constraint Generation}: Generate adjacency constraints based on CB terminal pairs, directional complementarity, and CP matching rules.
\end{enumerate}

\textbf{Implementation Approach}
The SoM-GI methodology enhances visual grounding by making implicit connections explicit through markers. For power system segments:

\textbf{Marker Visibility:}
Each type of marker employs unique visual encoding as follows:
\begin{itemize}
\item \textbf{Red squares}: Circuit breakers (active connections)
\item \textbf{Text labels}: Directional indicators (L\_i\_j\_direction)
\item \textbf{Blue markers}: CPs (boundary alignment)
\end{itemize}

\begin{figure*}[!h]
    \centering
    \begin{subfigure}{\textwidth}
        \centering
        \includegraphics[width=0.6\textwidth]{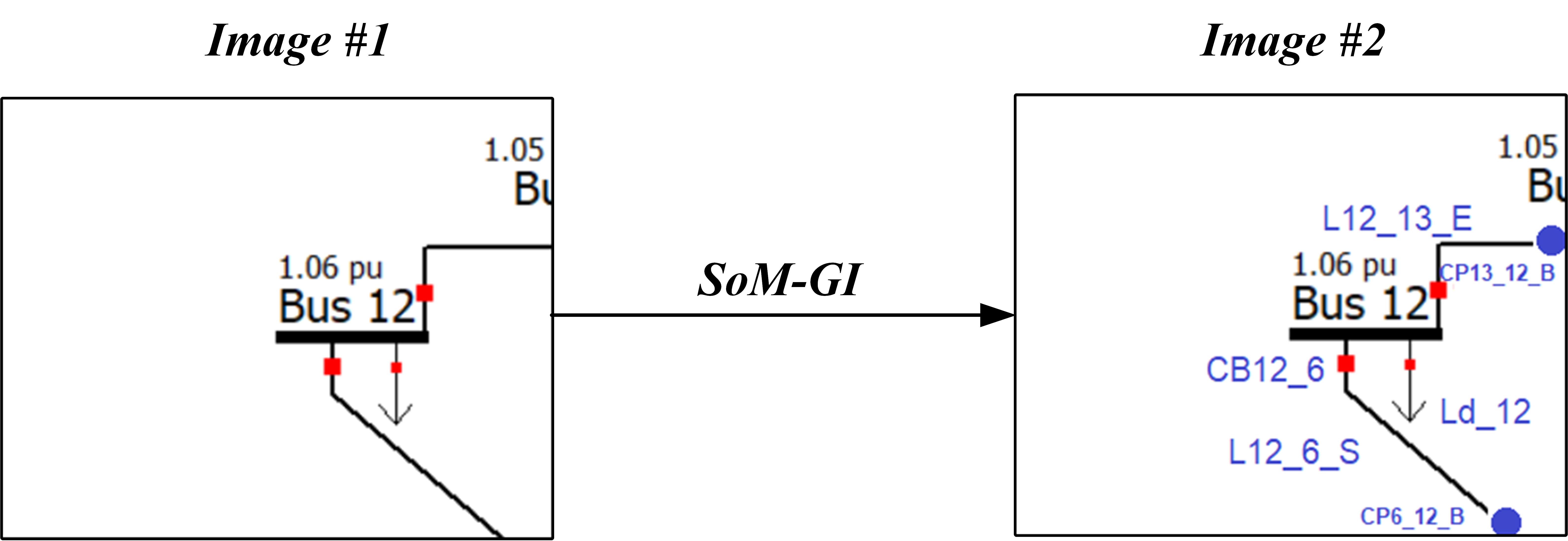}
        \label{fig:subfig1}
    \end{subfigure}
    
    \vspace{0.1cm} 
    
    \begin{subfigure}{\textwidth}
        \centering
        \small
        \begin{mdframed}[]
            Based on the comparison between the segment without SoM-GI (i.e., Image \#1) and with SoM-GI (i.e., Image \#2), here are the six blue indicators and their meanings:
            \begin{itemize}
                \item $\text{L}12\_13\_\text{E}$: The transmission line from Bus 12 to Bus 13 exits through the east direction of this segment.
                \item $\text{CP}13\_12\_\text{B}$: The point B for the Bus 13 to Bus 12 link, indicating this boundary point must connect to its matching $\text{CP}13\_12\_\text{A}$ in an adjacent segment.
                \item $\text{CB}12\_6$: CB identification between Bus 12 and Bus 6, marking the terminal equipment for this transmission line connection.
                \item $\text{Ld}\_12$: The load connected at Bus 12, indicating local power consumption at this bus.
                \item $\text{L}12\_6\_\text{S}$: The transmission line from Bus 12 to Bus 6 exits through the south direction of this segment.
                \item $\text{CP}6\_12\_\text{B}$: The point B for the Bus 6 to Bus 12 link, indicating this boundary point must connect to its matching $\text{CP}6\_12\_\text{A}$ in an adjacent segment.
       \end{itemize}
        \end{mdframed}
        \label{fig:subfig2}
    \end{subfigure}
    
    \caption{A sample of applied SoM-GI rules (i.e., blue marks) to a segment of an HMI screen.}
    \label{SoM-GI_rules}
\end{figure*}

\textbf{Spatial Reasoning:}
The arrangement leverages spatial relationships encoded in directional markers. For instance, $\text{L}1\_2\_\text{S}$ indicates bus 1 connects to bus 2 via the southern boundary, requiring the segment containing bus 2 to be placed south of the segment containing bus 1. A sample of an HMI screen segment with and without the SoM-GI method is represented in Fig.~\ref{SoM-GI_rules}. As can be seen, there are different indicators to enhance the understanding of the PF analysis in an HMI screen according to the transmission lines, CPs at edges/boundaries, CB markers, and load markers. The analogous SoM-GI methodology is applicable and can be extended to be utilized across a diverse range of sectors and segments, allowing for a comprehensive and interdisciplinary approach within various fields of study.

Finally, according to the provided information, the following recommendations/rules are crafted to train the GPT model in addition to the technical parts of PF analysis, as shown in the following box:
\begin{mdframed}[]
\textbf{Recommendations/Rules for the training process}
\small
\begin{itemize}
\item \textit{Red squares show the closed circuit breakers (CBs) at the terminals of each transmission line. For example, $\text{CB}1\_2$ and $\text{CB}2\_1$ should be at the terminals of one transmission line between bus 1 and bus 2. This rule applies to all buses with relevant CBs. Also, green squares show the open CBs so there is no flow of power. Also, it is not possible to have different statuses of CBs at both terminals of one single transmission line. That means that if we have a closed CB at one terminal, there should be another closed CB at another terminal to meet the normal scenario. Otherwise, this is an anomaly.}
\item \textit{Indicators ``$\text{L}i\_j\_\text{direction}$'' show the connection of different transmission lines in which N, S, W, E, NE, NW, SE, SW show North, South, West, East, Northeast, Northwest, Southeast, Southwest, respectively. These directions show the connection to the adjacent segment according to the transmission line connections. For instance, $\text{L}1\_2\_\text{S}$ (connection of bus 1 to bus 2 through south direction) and/or $\text{L}2\_1\_\text{N}$ show an adjacent of relevant segments in a proper way. This rule applies to all transmission lines and relevant buses.}
\item \textit{There are some CP at end of some parts for better clarification. Actually, CPs are boundary points in blue for better clarification. For instance, ``$\text{CP}1\_2\_\text{A}$'' at one edge and ``$\text{CP}1\_2\_\text{B}$'' at another edge. It means these two points should have connected each other. Also, in some case, there are ``$\text{CP}i\_j\_\text{C}$'' and ``$\text{CP}i\_j\_\text{D}$'' in which ``C'' and ``D'' should connect each other. That means that ``A'' should be connected to ``B'' and ``C'' should be connected to ``D''. Please check this rule carefully. Match connection pairs:}
\begin{itemize}
    \item \textit{A connects to B}
    \item \textit{C connects to D}
    \item \textit{Same $\text{CP}i\_j$ must connect}
    \item \textit{Arrange segments so matched CPs are adjacent.}
\end{itemize}
\end{itemize}
\end{mdframed}
This approach provides several advantages for power system visualization, including:
\begin{enumerate}
\item \textbf{Explicit Connectivity}: Visual markers make implicit connections explicit
\item \textbf{Systematic Arrangement}: Rule-based placement reduces ambiguity
\item \textbf{Verification}: Visual inspection confirms proper alignment
\end{enumerate}
Now, the goal is to implement the SoM-GI approach according to different segments, and check whether there is a correct rearrangement of different segments to satisfy various constraints, connections, and consistent visual information. 

The following ``Prompt''-``Response'' box shows that the SoM-GI approach can make a good connection between different segments in an HMI screen based on non-arranged parts. This framework provides a systematic approach that scales from $3 \times 3$ to any $N \times N$ grid while ensuring the integrity of electrical systems by satisfying various constraints imposed on them. The primary determinants of success involve prioritizing the alignment of transmission line orientations, strategically positioning hubs based on optimal connectivity indicators, and validating a comprehensive set of constraints to ensure system efficacy.

\begin{promptbox}
\small
\textbf{Scenario \#3A (SoM-GI)}
\vspace{2mm}

\noindent \textit{Arrange these 9 power system segments into a $3\times3$ grid. I need to trace the exact transmission line directions and connection points (CPs) systematically. Please check the rules and indicators to improve the capability of arrangement using the proposed methodology in this chat. Also, consider this arrangement and rules as ``Reference'' for future tasks.}
\end{promptbox}

\begin{figure*}[!h]
\centerline{\includegraphics[width=2.0\columnwidth]{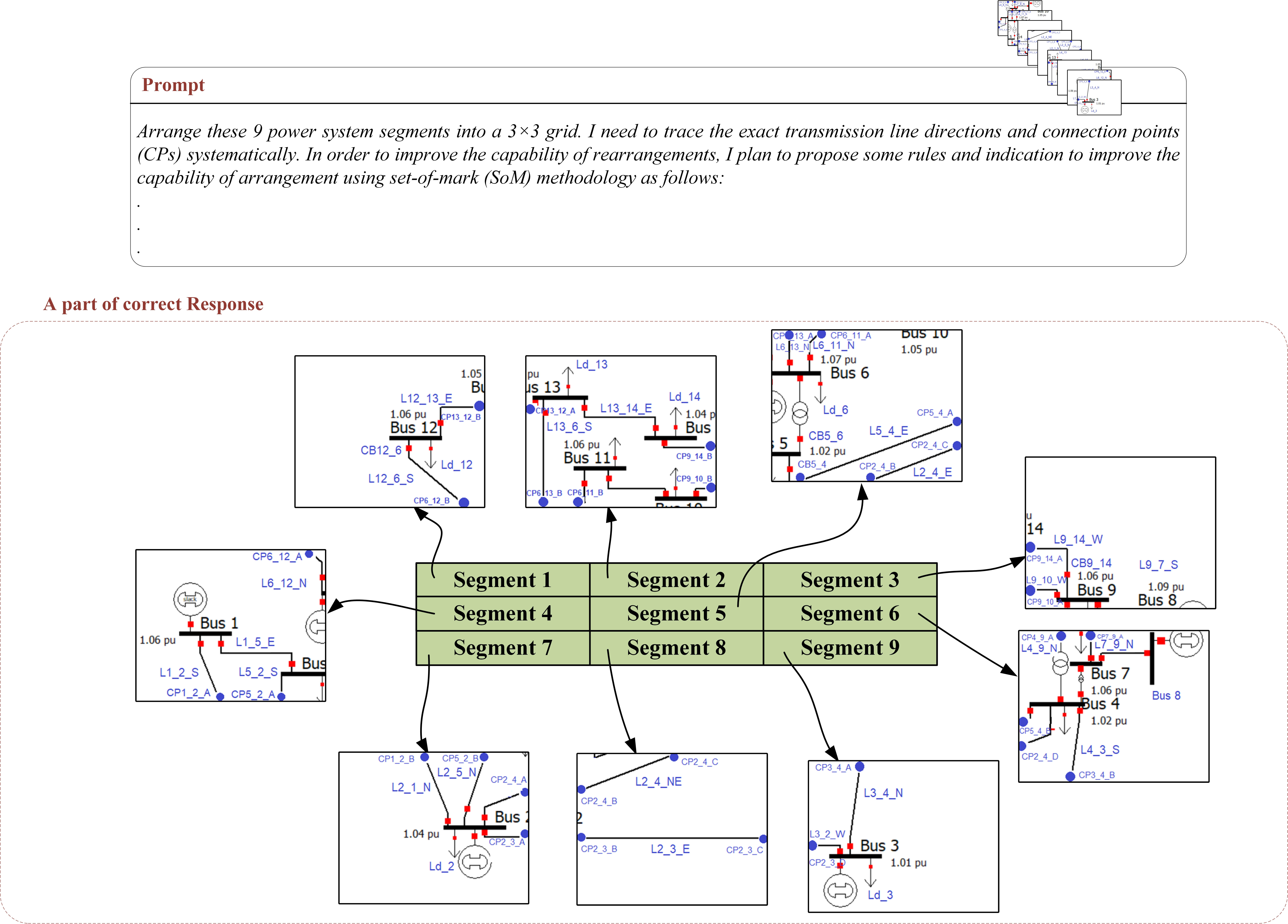}}
\caption{A correct rearrangement of different segments of a corrupted HMI screen using the SoM-GI Technique.}
\label{SoM-GI_fig_correct}
\end{figure*}

\begin{responsebox}
\small
\textbf{Scenario \#3A (SoM-GI)}
\vspace{2mm}

\noindent Looking at the 9 power system segments, I need to trace the transmission line connections and match the connection points (CPs) to arrange them in a $3\times3$ grid. Let me analyze the key connections:
\vspace{1.5mm}

\noindent \textbf{Transmission Line Directions:}
\begin{itemize}
    \item \texttt{L1\_2\_S} (Image 8): Bus 1 connects \textbf{south} to Bus 2
    \item \texttt{L2\_1\_N} (Image 5): Bus 2 connects \textbf{north} to Bus 1
    \item \texttt{L2\_3\_E} (Image 5): Bus 2 connects \textbf{east} to Bus 3
    \item \texttt{L3\_2\_W} (Image 9): Bus 3 connects \textbf{west} to Bus 2
    \item \texttt{L3\_4\_N} (Image 9): Bus 4 connects \textbf{north} from Bus 3
    \item \texttt{L4\_3\_S} (Image 1): Bus 4 connects \textbf{south} to Bus 3
    \item \texttt{L6\_12\_N} (Image 8): Bus 6 connects \textbf{north} to Bus 12
    \item \texttt{L12\_6\_S} (Image 7): Bus 12 connects \textbf{south} to Bus 6
    \item \texttt{L12\_13\_E} (Image 7): Bus 12 connects \textbf{east} to Bus 13
    \item \texttt{L13\_6\_S} (Image 6): Bus 13 connects \textbf{south} to Bus 6
\end{itemize}
\textbf{Connection Point Matching:}
\begin{itemize}
    \item \texttt{CP1\_2\_A} (Image 8) $\leftrightarrow$ \texttt{CP1\_2\_B} (Image 5)
    \item \texttt{CP2\_3\_A} (Image 5) $\leftrightarrow$ \texttt{CP2\_3\_B} (Image 4)
    \item \texttt{CP2\_3\_C} (Image 4) $\leftrightarrow$ \texttt{CP2\_3\_D} (Image 9)
    \item \texttt{CP3\_4\_A} (Image 9) $\leftrightarrow$ \texttt{CP3\_4\_B} (Image 1)
    \item \texttt{CP6\_12\_A} (Image 8) $\leftrightarrow$ \texttt{CP6\_12\_B} (Image 7)
    \item \texttt{CP13\_12\_A} (Image 6) $\leftrightarrow$ \texttt{CP13\_12\_B} (Image 7)
\end{itemize}

\noindent Based on these connections, the \textbf{3\(\times\)3} grid arrangement should be:
\begin{center}
\begin{tabular}{|c|c|c|}
\hline
\textbf{Bus 12} & \textbf{Bus 13, 11, 10} & \textbf{Bus 14, 9, 8} \\
\hline
\textbf{Bus 1, 5} & \textbf{Bus 10, 6, 5} & \textbf{Bus 7, 4, 8} \\
\hline
\textbf{Bus 2} & \textbf{Connection diagram} & \textbf{Bus 3} \\
\hline
\end{tabular}
\end{center}

\noindent This arrangement ensures:
\begin{itemize}
    \item All connections align properly (e.g., Bus 1 above Bus 2).
    \item All connection points (CPs) match between adjacent segments.
    \item Circuit breakers at transmission line terminals are properly connected.
\end{itemize}
\end{responsebox}
According to these results, it can be seen from Fig.~\ref{SoM-GI_fig_correct} that the SoM-GI technique can arrange different segments in correct positions considering its semantic understanding. 
This GenAI-based methodology uses guidelines based on markers and indicators to improve the capability of the rearrangement process. Compared with the GenAI-based AD without the SoM (e.g., markers, indicators, and rules), the GPT model solely has no idea about the connections, rules, and the power system domain.
\subsubsection{Scenario \#3B: A CB Malfunction Between Bus 6 and Bus 13}
\begin{figure}[!h]
\centerline{\includegraphics[width=0.8\columnwidth]{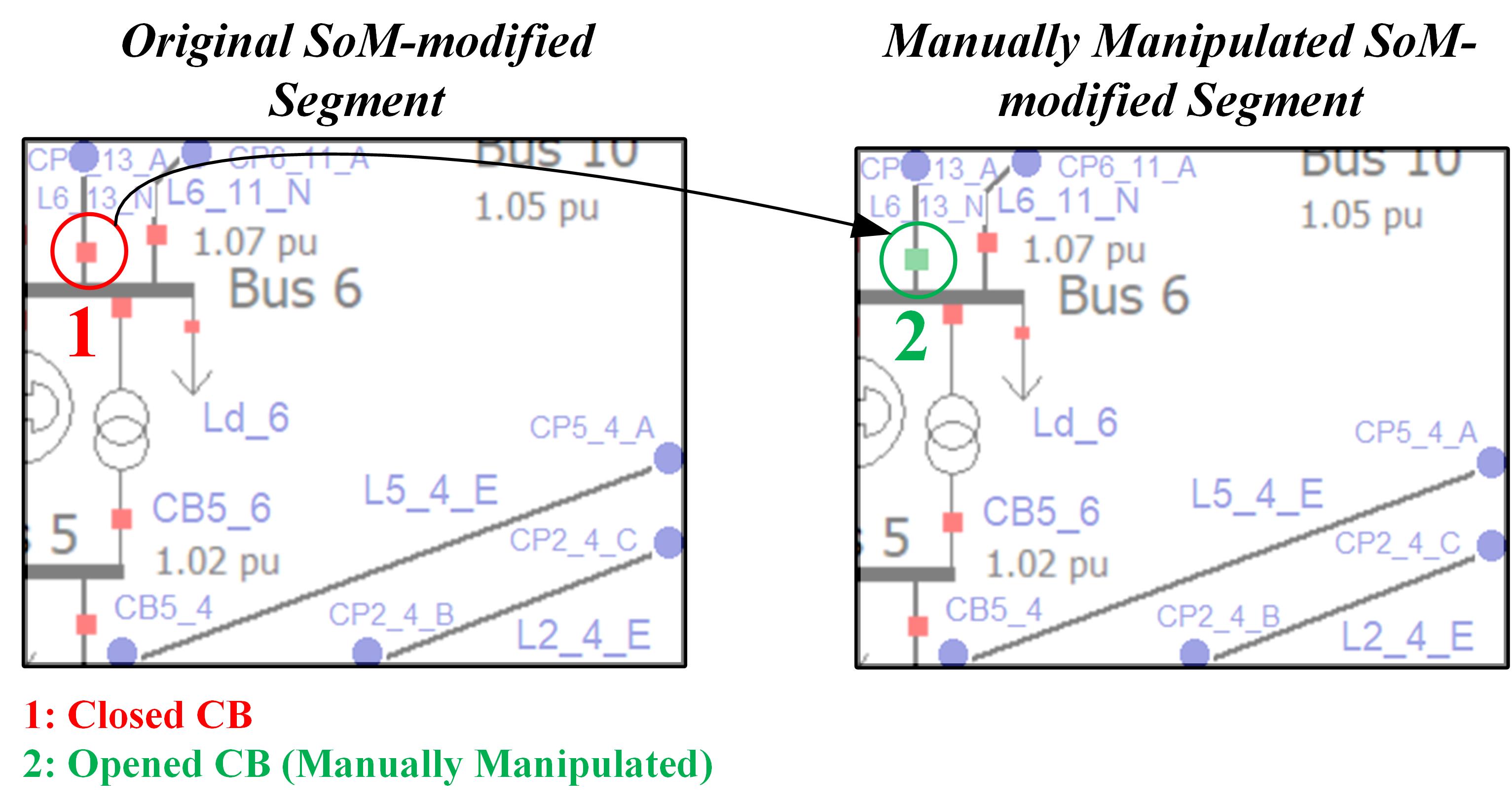}}
\caption{A manual manipulation of CB between Bus 6 and Bus 13 to an open status.}
\label{SoM-GI_fig_CB6_13_anomaly}
\end{figure}

In addition, to show the capability of the proposed SoM-GI technique, a DI is applied to a part of a segment. Then, a prompt is given to request potential anomalies in the uploaded segments and perform the rearrangement process simultaneously. Hence, one of the CBs between Bus 6 and Bus 13 (i.e., $\text{CB}\_6\_13$) is manually manipulated to an open status (i.e., green color) - as shown in Fig.~\ref{SoM-GI_fig_CB6_13_anomaly}, to check whether the proposed method can detect this anomaly.
The ``Prompt'' and ``Response'' of the given scenario are mentioned regarding the reference box. 
\begin{promptbox}
\small
\textbf{Scenario \#3B (SoM-GI)} 
\vspace{2mm}

\noindent \textit{Now, the purpose is to do an anomaly detection process. Can you find any anomaly compared with ``Reference''? If yes, please specify the reason.}
\end{promptbox}

\begin{responsebox}
\small
\textbf{Scenario \#3B (SoM-GI)}
\vspace{2mm}

\noindent Looking at the images carefully and comparing them with the reference configuration, I found \textbf{one anomaly}:
\paragraph{Anomaly Detected:} In Segment \#5, there is a GREEN circuit breaker at \text{CB6\_13} (the connection between Bus 6 and Bus 13).

\paragraph{Reason for Anomaly:} According to the rules established in the reference:
\begin{itemize}
    \item In normal operating conditions, all circuit breakers should be RED (closed) to allow power flow.
    \item Both terminals of a transmission line must have matching CB status.
\end{itemize}
However, in Segment \#5:
\begin{itemize}
    \item \text{CB6\_13} appears as GREEN (open), which means no power can flow through this line.
    \item This creates an abnormal condition where the transmission line between Bus 6 and Bus 13 is disconnected.
\end{itemize}
This violates the normal scenario conditions where all transmission lines should be in service with closed (red) circuit breakers at both terminals. The open CB at this location would interrupt the power flow between Bus 6 and Bus 13, potentially affecting system reliability and power delivery. All other aspects (direction indicators, connection points, and other CB statuses) appear to match the reference configuration correctly.

\end{responsebox}
It is apparent that this SoM-GI method can detect the abnormal scenario of manual opening of the CB between Bus 6 and Bus 13. Also, it interpreted the reference model along with all rules, indicators, and markers and made a visual comparison considering the SE and PF rules. 
\subsubsection{Scenario \#3C: False Data Injection at Bus 2}
In this part, an FDI is applied to one of the segments (i.e., changing the voltage magnitude of Bus 2 from 1.04 pu to 1.02 pu) to check with the SoM-GI technique whether it can detect this anomaly considering the rules, as presented in the following interaction:

\begin{promptbox}
\small
\textbf{Scenario \#3C (SoM-GI)} 
\vspace{2mm}

\noindent \textit{The purpose is to do an anomaly detection process again. Can you find any anomaly compared with ``Reference''? If yes, please specify the reason.}
\end{promptbox}
\begin{responsebox}
\small
\textbf{Scenario \#3C (SoM-GI)}
\vspace{2mm}

\noindent Looking at the images carefully and comparing them with the reference configuration, I found \textbf{one anomaly:}

\paragraph{Anomaly Detected:} In Segment \#7, Bus 2 shows a voltage of $1.02$ pu instead of the normal $1.04$ pu.

\paragraph{Reason for Anomaly:} According to the reference configuration:
\begin{itemize}
    \item Bus 2 should have a voltage magnitude of $1.04$ pu in normal operating conditions.
    \item However, in Segment \#7, Bus 2 displays $1.02$ pu.
\end{itemize}

\noindent This represents a voltage magnitude anomaly where Bus 2 is experiencing lower than normal voltage. This could indicate:
\begin{itemize}
    \item A voltage regulation issue
    \item Increased loading conditions causing voltage drop
    \item A state estimation error
    \item Potential power flow convergence issues
\end{itemize}
\noindent All other aspects appear normal:
\begin{itemize}
    \item All circuit breakers are red (closed) as expected.
    \item Direction indicators match properly.
    \item Connection points align correctly with the reference arrangement.
    \item Other bus voltages match their reference values.
\end{itemize}
\noindent This voltage deviation at Bus 2 could impact system stability and power quality, as it represents approximately a $2\%$ drop from the normal operating voltage.
\end{responsebox}
The proposed method proved remarkably effective in detecting subtle power system anomalies that could easily be missed during operations. In this test scenario, the system successfully identified a single but significant voltage deviation at Bus 2, where it detected a 1.02 p.u. reading instead of the expected 1.04 p.u. – an apparently small $2\%$ drop that could signal serious impacts (e.g., voltage regulation problems, overloading conditions, or measurement errors). While correctly identifying this anomaly, it simultaneously verified that all other system components were operating normally -- CBs were properly closed, power flow directions were correct, and remaining bus voltages matched their reference values. This demonstrates the SoM-GI approach's ability to act similarly to an expert operator who not only detects problems but understands their context, providing valuable insights about potential causes ranging from increased system loading to SE errors, finally helping operators make informed decisions to maintain grid stability and power quality.
\subsection{LIMITATIONS AND CONSIDERATIONS}
The GenAI-based framework proposed for enhancing cybersecurity in EMSs demonstrates significant potential, yet it is crucial to recognize several key limitations for a comprehensive assessment. Initially, the existing system relies on cloud-based API interactions with Anthropic Claude Pro, causing delays in response times. These delays are inadequate for real-time control tasks requiring responses within sub-second intervals. To overcome this constraint, it is probable that edge-computing systems or locally deployed AI models will need to be utilized to satisfy demanding temporal constraints. 

Secondly, the framework's reliance on external cloud services could present vulnerabilities, such as dependence on service availability, risk of sensitive grid data exposure, ongoing subscription expenses, and restricted authority over model updates or retraining timelines. Thirdly, the system's performance is significantly dependent on the quality, quantity, and diversity of its training data. Despite the fact that validation was conducted utilizing different scenarios on the IEEE 14-bus test system, additional comprehensive evaluation is imperative to establish both scalability and robustness of the approach, particularly when it is deployed on larger grid systems or when encountering unforeseen attack scenarios that were not part of the initial validation set. Models optimized for one setup may fail to generalize well to different configurations without re-optimization or retraining. As systems expand, scalability becomes a significant challenge, especially with the increasing incorporation of distributed energy resources (DERs). The SoM-GI methodology may experience exponential computational complexity with larger HMI displays, and the processing of high-frequency PMU data streams will require additional algorithmic optimization. In addition, while GenAI offers understandable explanations in natural language, its somewhat ambiguous reasoning can lead to trust issues for operators familiar with deterministic approaches. This matter is especially crucial in safety-critical settings, where AI tools should function as decision-support mechanisms rather than independent decision entities. A significant challenge lies in regulatory compliance. The existing NERC CIP standards provide insufficient direction for the certification, validation, and auditing processes of AI-based systems, potentially hindering their deployment in regulated utility contexts. The validation performed with simulated IEEE 14-bus data, though controlled and systematic, fails to entirely encompass the complexities found in real-world power systems. These complexities include factors such as sensor noise, communication delays, and measurement uncertainties. Consequently, there is a need for future field testing under actual operational conditions. Ultimately, the costs associated with implementation including computing infrastructure, software licenses, staff training, and maintenance must be thoroughly evaluated. These expenses could lead to discrepancies, as only large utilities might be able to afford sophisticated AI-based defense systems. Despite these limitations, the proposed GenAI-based framework marks a significant step forward in the cybersecurity of modern power systems. When used as an additional layer in a defense-in-depth strategy, it has the potential to significantly enhance traditional protection systems and human expertise, leading to more robust and adaptive grid security.
\section{Conclusions and Future Directions} \label{conclusion-section}
This study has effectively developed an innovative security framework designed to secure power grid control systems from advanced threats/errors targeting multiple points of vulnerability. By integrating state-of-the-art AI technology with a comprehensive understanding of power system operations, three pioneering solutions are proposed including a GenAI-based ADS capable of identifying stealth attacks that conventional BDD (e.g., $\chi^2$ test) miss, a GenAI-based ADS for identifying FDIs by training the GPT model with PF results and SE rules, and the SoM-GI approach, which enables AI to interpret and comprehend power grid displays comparable to the expertise of experienced operators. Empirical evaluation utilizing actual power system models has demonstrated the framework's proficiency in identifying a range of threats, including deceptive data manipulations that align mathematically yet violate physical laws, and visual manipulations on HMI screens that might mislead human operators into making risky decisions. The distinguishing feature of this approach is its capacity to function similarly to both an engineer and a detective, synthesizing evidence from numerical data, visual inputs, and system behavior patterns to detect anomalies escaping traditional security mechanisms, while also conveying findings in a way that operators can easily interpret and trust.

In future research, the enhancement of power grid security will be achieved through a series of interconnected initiatives designed to tackle the evolving challenges posed by technology advancements. Firstly, it is imperative to create robust frameworks for validation that leverage adversarial training methodologies to verify AI-generated outputs, ensuring alignment with a wide array of operational specifications and physical constraints. Concurrently, the development of sophisticated real-time analytical systems is crucial for processing streaming data from HMI with a latency measured in sub-seconds, all while maintaining computational efficiency via optimal resource allocation. Furthermore, the framework will evolve to integrate novel DERs, microgrids, and virtual power plants, each of which fundamentally transforms conventional grid operational dynamics. Significant attention will be directed towards the design of advanced detection algorithms that can recognize extensive coordinated attacks targeting specific topological vulnerabilities within the network, especially those affecting zero-flow transmission segments that are not detectable by standard statistical monitoring techniques. In addition, user interface systems will be progressively adaptive, with the ability to modify the presentation of information based on both the severity of threats and the expertise level of operators. This research will also explore the development of proactive defense strategies against AI-led cyber threats by implementing continuous adversarial learning processes. Finally, a groundwork will be considered for security protocols resistant to quantum-based attacks, ensuring the sustainable protection of infrastructure as the industry transitions to cryptographic paradigms designed for a post-quantum world.
\bibliographystyle{IEEEtran}
\bibliography{IEEEabrv,RefDatabase}

\begin{IEEEbiography}[{\includegraphics[width=1.0in,height=1.5in,clip,keepaspectratio]{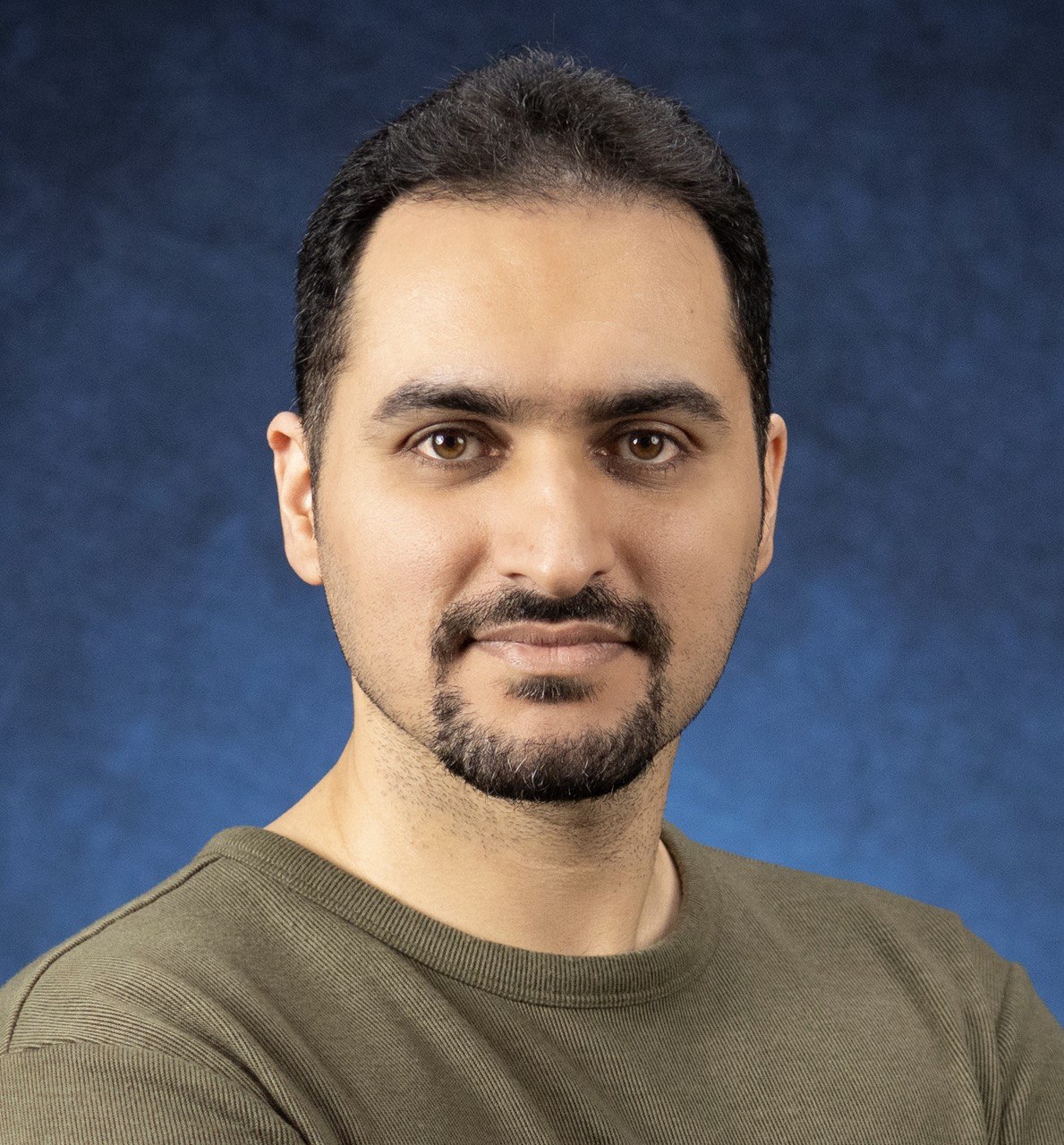}}]
{Aydin Zaboli} (Member, IEEE) received the Ph.D. degree in electrical, electronics, and computer engineering from the University of Michigan, MI, USA. His research interests include smart grid security, autonomous vehicles, anomaly detection, energy management systems, SCADA, transportation electrification, renewable energy resources, and load forecasting. He was a recipient of the Rackham Predoctoral Fellowship from the University of Michigan-Rackham Graduate School, from 2024 to 2025. He has served as a reviewer for more than 250 papers in prestigious journals and conferences, particularly IEEE Transactions on Smart Grids, IEEE Transactions on Transportation Electrification, IEEE Access, IEEE Transactions on Vehicular Technology, contributing to the advancement of research in smart grids and transportation electrification.
\end{IEEEbiography}


\begin{IEEEbiography}[{\includegraphics[width=1in,height=1.25in,clip,keepaspectratio]{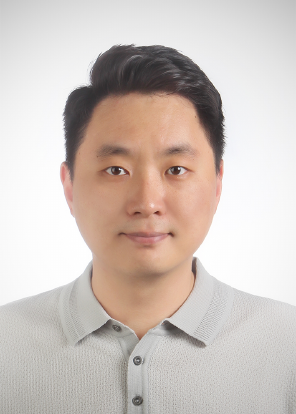}}]{Junho Hong} (Senior Member, IEEE) is an Associate Professor in the Department of Electrical and Computer Engineering at the University of Michigan–Dearborn. He received his Ph.D. degree in Electrical Engineering, specializing in the cybersecurity of substation automation systems, from Washington State University, Pullman, in 2014. From 2014 to 2019, he worked at ABB, where he provided technical project leadership and supported strategic corporate technology development and productization in areas related to cyber-physical security for substations, power grid control and protection, renewable integration, and utility communications. He has since been engaged in research on the cybersecurity of energy delivery systems with the U.S. Department of Energy (DOE), serving as Principal Investigator (PI) and Co-PI on projects involving substations, microgrids, HVDC, FACTS, and high-power EV chargers. He holds 14 U.S. patents and serves as a U.S. Regular Member of CIGRE Study Committee D2.
\end{IEEEbiography}

\begin{IEEEbiography}[{\includegraphics[width=1in,height=1.25in,clip,keepaspectratio]{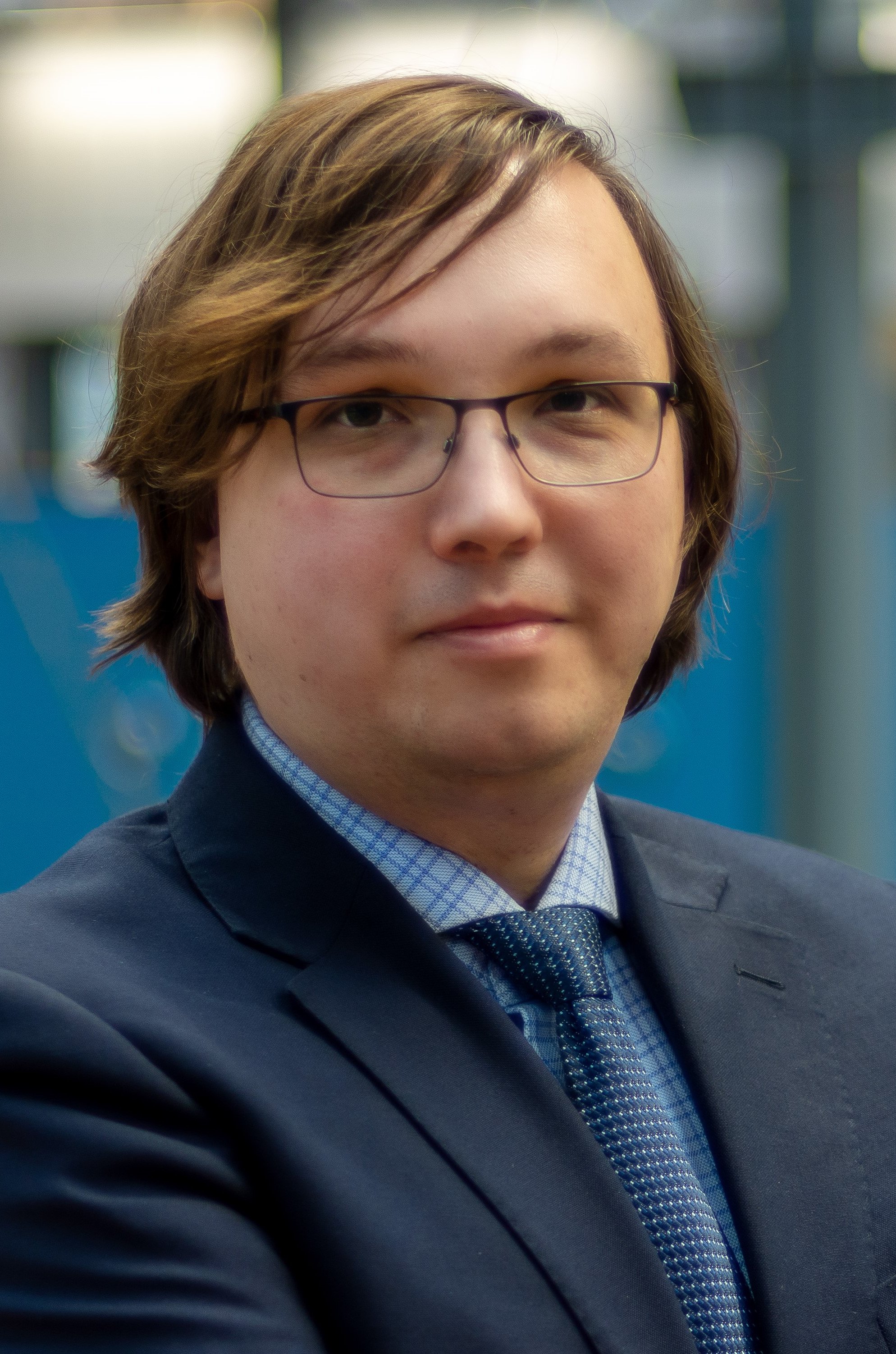}}]{Alexandru \c{S}tefanov} (Member, IEEE) received the M.Sc. degree from the University Politehnica of Bucharest, Romania, in 2011, and the Ph.D. degree from University College Dublin, Ireland, in 2015. He is an Associate Professor in intelligent electrical power grids in the Department of Electrical Sustainable Energy at TU Delft, The Netherlands. He is the Director of the Control Room of the Future (CRoF) Technology Centre. He is leading the Cyber Resilient Power Grids (CRPG) research group. His research interests include cyber security of power grids, resilience of cyber-physical systems, and next generation grid operation. He holds the professional title of Chartered Engineer from Engineers Ireland.
\end{IEEEbiography}

\begin{IEEEbiography}[{\includegraphics[width=1in,height=1.25in,clip,keepaspectratio]{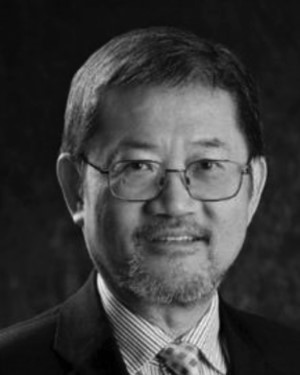}}]{Chen-Ching Liu} (Life Fellow, IEEE) received the Ph.D. degree from the University of California at Berkeley, Berkeley, CA, USA, in 1983. He is currently an American Electric Power Professor of Electrical Engineering at Virginia Tech. He is also a member of the Virginia Academy of Science, Engineering, and Medicine, and the U.S. National Academy of Engineering. He was a recipient of the IEEE Third Millennium Medal in 2000 and the IEEE Power and Energy Society Outstanding Power Engineering Educator Award in 2004.
\end{IEEEbiography}

\begin{IEEEbiography}[{\includegraphics[width=1in,height=1.25in,clip,keepaspectratio]{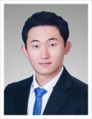}}]{Chul-Sang Hwang} (Member, IEEE) received the B.S. and Ph.D. degrees in electrical engineering from Changwon National University, Changwon, Korea, in 2010 and 2016, respectively. He worked for Korea Electrotechnology Research Institute (KERI) as a Research Assistant from 2017 to 2018. He joined Korea Electrotechnology Research Institute (KERI) in 2018, where he is currently a Senior Research Associate in the Smart Grid Research Division System Reliability Research Team. His research interests include distribution network, battery management system, microgrid, and performance evaluation technology.
\end{IEEEbiography}

\EOD

\end{document}